\documentclass[aps,twocolumn,floatfix,prd,superscriptaddress,nofootinbib]{revtex4}
\usepackage[dvips]{graphicx}
\usepackage{amsmath}
\usepackage{amssymb}
\usepackage{bbm}
\newcommand{\beq}{\begin{equation}}
\newcommand{\eeq}{\end{equation}}
\newcommand{\bea}{\begin{eqnarray}}
\newcommand{\eea}{\end{eqnarray}}
\newcommand{\Z}{\mathbbm{Z}}
\newcommand{\ave}[1]{\langle {#1} \rangle}
\newcommand{\eq}[1]{Eq.~(\ref{#1})}
\newcommand{\eqs}[1]{Eqs.~(\ref{#1})}

\renewcommand{\=}{\;=\;}

\def\pslash{p\hspace{-1.7mm}/}
\def\pvslash{\vec p\hspace{-1.7mm}/}
\def\phslash{\hat p\hspace{-1.7mm}/}

\def\dslash{\partial\hspace{-2.0mm}/}
\def\muslash{\mu\hspace{-2.0mm}/}
\def\unity{1\hspace{-1.4mm}1}

\def\={\;=\;}
\def\+{\;+\;}
\def\bear{\begin{array}}
\def\ear{\end{array}}
\def\Tr{\mathrm{Tr}}

\begin{document}

\title{\hspace*{\fill}{\small\sf MIT-CTP 3964}\\[4mm]Solitonic ground states in (color-) superconductivity}


\author{Dominik Nickel}
\affiliation{Center for Theoretical Physics, MIT, 
Cambridge, MA 02139, USA
}

\author{Michael Buballa}
\affiliation{Institut f\"ur Kernphysik, Technische Universit\"at Darmstadt, Germany}


\date{\today}


\begin{abstract}

We present a general framework for analyzing inhomogeneous (color-) 
superconducting phases in mean-field approximation without restriction
to the Ginzburg-Landau approach. 
As a first application, we calculate real gap functions with general 
one-dimensional periodic structures for a $3+1$-dimensional 
toy model having two fermion species. 
The resulting solutions are energetically favored against homogeneous 
superconducting (BCS) and normal conducting phases in a window for the chemical potential difference $\delta\mu$
which is about twice as wide as for the most simple plane-wave ansatz 
(``Fulde-Ferrell phase'').
At the lower end of this window, we observe the formation of a soliton
lattice and a continuous phase transition to the BCS phase.
At the higher end of the window the gap functions are sinusoidal, and
the transition to the normal conducting phase is of first order.
We also discuss the quasiparticle excitation spectrum in the
inhomogeneous phase.
Finally, we compare the gap functions with the known analytical solutions
of the $1+1$-dimensional theory.

\end{abstract}


\maketitle


\section{Introduction}
Inhomogeneous ground states due to imbalanced Fermi surfaces have been
discussed in various contexts.
Theoretical investigations started off by considering a clean paramagnetic superconductor exposed to an
external magnetic field.
For such a system Fulde and Ferrell analyzed the ground state with the order
parameter, i.e., the gap function, forming a plane wave~\cite{Fulde:1964zz}.
Larkin and Ovchinnikov extended their work by considering more general
inhomogeneous ground states, but relying on the Ginzburg-Landau
expansion~\cite{LO64}.

In recent years these ideas have attracted new interest from two fields. 
One of them are systems of ultra-cold atoms~\cite{cold-atoms-review},
where new fascinating techniques open unprecedented possibilities to 
study the pairing of imbalanced Fermi systems in a trap. 
Here, unlike in solids where the electron interaction is often difficult do understand in detail, imbalanced systems can be prepared in a rather straight forward and controlled way.

In this paper, we will mainly aim at color superconductivity in
deconfined quark matter. Here the problem of imbalanced Fermi 
surfaces is almost unavoidable. 
It is expected that color superconducting phases are present in
the QCD phase diagram at sufficiently high densities and low 
temperatures (see Refs.~\cite{Rajagopal:2000wf,Alford:2001dt,Schafer:2003vz,
Rischke:2003mt,Buballa:2003qv,Shovkovy:2004me,Alford:2007xm} 
for corresponding reviews).
In nature, the most promising places to find these conditions 
are the centers of neutron stars.
Here the system must be in beta equilibrium and, at least globally,
electrically and color neutral.
This would be fulfilled if there were equally many up-, down-, 
and strange quarks.
However, at densities which can be reached in neutron stars,
strange quarks are expected to be considerably suppressed by their mass. 
In turn, this forces the density of down quarks to be larger than
the density of up quarks in order to achieve electric neutrality.
Since, on the other hand, the most attractive channels involve 
quarks of unequal flavors, we are naturally led to the problem
of pairing in an imbalanced Fermi system
\cite{Alford:1999pa,Bedaque:1999nu,Buballa:2001gj,Alford:2002kj,Steiner:2002gx,
Nickel:2006kc,Ruster:2004eg,Ruster:2005jc,Blaschke:2005uj,Abuki:2005ms,
Nickel:2008ef}.

Let us briefly recall what the problem actually is. 
In BCS theory, pairing occurs among fermions with opposite momenta,
forming Cooper pairs with zero total momentum.
If both fermions are at their respective Fermi surface,
the pair can be created at no free-energy cost and the pairing is
is always favored as soon as there is an attractive interaction.
This is, however, no longer the case if the Fermi momenta of the 
fermions to be paired are unequal.
BCS pairing then requires that the Fermi spheres first have to be 
equalized.
In the case of quark matter this could be realized, e.g., in a weak 
process which replaces some of the down quarks by strange quarks.
Of course, this will only be favorable if the free energy which is
needed for this process is overcompensated by the pairing energy.
This sets a limit for this mechanism in terms of the Fermi momentum 
difference in the unpaired system and the BCS gap \cite{ClCh1962}.

Therefore the question arises how the system reacts if the imbalance
does no longer allow for BCS-like pairing. 
Sticking to homogeneous phases, some authors have suggested so-called
gapless or breached pairing phases \cite{Sarma,ShHu,breached,gCFL},
where equal Fermi surfaces are created by lifting some of the 
fermions to higher momentum states. 
At these ``new'' Fermi surfaces the fermions can again form Cooper pairs
with zero total momentum.
It was found, however, that this pairing mechanism suffers from 
instabilities \cite{Huang:2004bg,Casalbuoni:2004tbg}.
In atomic systems this will most likely lead to a phase separation
into a BCS-like phase with equal densities and an unpaired phase with 
unequal densities. 
In principle, something similar could happen in quark matter as well
\cite{Bedaque:1999nu,Neumann:2002jm,Reddy:2004my}. 
However, because of long-range Coulomb forces, the different
phase domains cannot grow arbitrarily large, and it is therefore
unclear whether a mixed phase can exist at all.

Instead, it seems reasonable that the matter becomes inhomogeneous
already on a microscopic scale by the formation of ``crystalline
condensates'' (see Ref.~\cite{Casalbuoni:2003wh} for a dedicated review).
The basic idea is to form Cooper pairs with non-zero total momentum.
This has the obvious advantage that the fermions in the pair no longer
have to have opposite momenta, and therefore each of them can stay
on its respective Fermi surface.
In the context of color superconductors, this possibility has been
investigated first in Ref.~\cite{Alford:2000ze}.  
The authors restricted themselves to a two-flavor model with a
plane-wave ansatz for the gap function, like in the 
original work by Fulde and Ferrell~\cite{Fulde:1964zz}. 
Indeed, as was shown in Ref.~\cite{GiannakisRen}, 
one of the instabilities which 
occur in gapless two-flavor color superconductors could be related to 
an instability against the formation of a Fulde-Ferrell (FF) -like 
condensate. 

On the other hand, since in the FF ansatz the total momentum of the 
pair is restricted to a non-zero but constant value $2\vec q$, 
this pairing pattern is strongly disfavored by phase space in most 
cases.
Several authors have therefore extended the ansatz to multiple
plane waves, studying both, two- and three-flavor 
systems~\cite{Bowers:2002xr,Casalbuoni:2005zp,Mannarelli:2006fy,Rajagopal:2006ig}.
As expected, the resulting solutions were found to be strongly favored
against the FF phase. 
However, these analyses were restricted to a Ginzburg-Landau 
approximation.
This turned out to be especially problematic in the two-flavor case,
where the Ginzburg-Landau functional for the energetically favored 
solution was not bounded from below \cite{Bowers:2002xr}.
Moreover, the crystal structures considered so far have been restricted
to superpositions of a finite number of plane waves whose wave vectors
all have the same length, whereas one should also allow for the
superposition of different wave lengths.

The aim of this paper is to overcome the restriction to
the Ginzburg-Landau approximation 
and to approach the mean-field problem explicitly.
The thermodynamic potential
is then always bounded from below and a proper treatment leads
to new insights which could not be obtained in the previous investigations.
As a first step we will focus on two-flavor pairing allowing an arbitrary
real gap functions with general one-dimensional periodic structure.

For inhomogeneous ground states the mean-field problem is already non-trivial
and requires to solve the Bogoliubov-de Gennes equations~\cite{deGennes66}.
Only for $1+1$-dimensional systems there is a good understanding of the
mean-field ground state and the thermodynamic properties of the
system~\cite{peierls1955,dashen1975,shei1976,horovitz1981,Thies,machida1984,Basar:2008im}.
This is, however, lacking for higher dimensional systems and attempts
have been made to simplify these equations, e.g., by integrating out
short-range fluctuations~\cite{Eilenberger68,LO68}.
We will not pursue such a direction here, but instead present a numerical
approach to solve the Bogoliubov-de Gennes equations in a convenient basis.
The presentation and derivation is elementary so that no prior knowledge of
inhomogeneous phases is required.
It turns out that at least for relativistic systems the
regularization of the theory has to be addressed carefully in order to avoid
undesired artifacts.

The paper is organized as follows:
In section~\ref{sec:formalism} we introduce the model we aim to investigate in
a certain approximation and
derive an expression for the thermodynamic potential in an inhomogeneous phase together with the corresponding gap equation.
Because of its importance for inhomogeneous phases in $3+1$-dimensions, we 
also discuss a suitable regularization scheme.
In section~\ref{sec:results} we then present numerical results for
inhomogeneous phases with one-dimensional inhomogeneity in $3+1$-dimensions.
As a prelude we discuss the homogenous (color-)superconducting and the
Fulde-Ferrell phase first, before confronting them with results for a general
pairing pattern.
For the latter we continue by exploring the quasi-particle spectrum and a
comparison to analytical results obtained in $1+1$-dimensions.
Finally we summarize our results in section~\ref{sec:summary} and give an
outlook for possible further investigations.


\section{Formalism}
\label{sec:formalism}

In this section we develop the general framework for the description
of inhomogeneous color-superconducting phases.

\subsection{Model Lagrangian}
\label{Lagrangian}

We consider an NJL-type
Lagrangian for massless quarks $q$ with three flavor 
and three color degrees of freedom,
\beq
    \mathcal{L} = \bar q \,(i\dslash + \muslash)\,q + 
    \mathcal{L}_\mathit{int}\,.
\eeq
We have introduced the notation $\muslash = \mu\gamma^0$,
where $\mu$ is the chemical potential. To be precise, $\mu$ is a diagonal 
matrix in color-flavor space, allowing for different chemical potentials for different colors 
or flavors. 

The interaction term is given by
\beq
    \mathcal{L}_\mathit{int} 
    = H \hspace{-3mm} \sum_{A,A'=2,5,7} \hspace{-3mm}
      ( \bar q \,i \gamma_5 \tau_A \lambda_{A'}\, q_C ) 
      ( \bar q_C \,i \gamma_5 \tau_A \lambda_{A'}\,q  )\,. 
\label{Lint}
\eeq
Here $H$ is a dimensionful coupling constant and 
$q_C(x) = C\bar q^T(x)$, where
$C=i \gamma^2 \gamma^0$ is the matrix of charge conjugation. 
$\tau_A$ and $\lambda_{A'}$ denote the antisymmetric Gell-Mann matrices 
acting in flavor space and color space, respectively. 
Thus, $\mathcal{L}_\mathit{int}$ corresponds to a quark-quark interaction
in the scalar flavor-antitriplet color antitriplet channel.

The above Lagrangian should be viewed as a typical example which allows
for the most important pairing patterns in color superconductivity,
like the two-flavor superconducting (2SC) phase and the 
color-flavor locked (CFL) phase. However, the formalism
we are going to develop in this section is by no means restricted to this 
model. In particular we may add mass terms, and the inclusion of
other interaction channels is straight forward. 

Applying standard bosonization techniques, the interaction term,
\eq{Lint}, can equivalently be rewritten as
\begin{alignat}{1}
    \mathcal{L}_\mathit{int} 
    = \frac{1}{2}
    \sum_{A,A'} \Big\{\;
      &(\bar{q}\, \gamma_5 \tau_A \lambda_{A'}\,q_C)\;\varphi_{AA'} 
\nonumber \\ 
      -\,&\varphi_{AA'}^\dagger\;
      (\bar q_C\, \gamma_5 \tau_A \lambda_{A'}\,q)
      \;-\, \frac{ 1}{2H}\, \varphi_{AA'}^\dagger\,\varphi_{AA'} \Big\},
\label{Lintbos}
\end{alignat}
with the auxiliary complex boson fields $\varphi_{AA'}(x)$, 
which, by the equations of motion,
\begin{alignat}{2}
    \varphi_{AA'}(x) \,&=\,& -2H&\,\bar q_C(x)\,\gamma_5 \tau_A \lambda_{A'}
                                  \,q(x)\,, 
\nonumber \\
    \varphi^\dagger_{AA'}(x) \,&=\,& 2H&\,
    \bar q(x)\, \gamma_5 \tau_A \lambda_{A'} \,q_C(x)\,, 
\end{alignat}
can be identified with scalar diquarks.

In mean field approximation we replace these quantum fields by their
expectation values
\begin{alignat}{1}
    \ave{\,\varphi_{AA'}(x)} &= \Delta_A(x)\,\delta_{AA'}\,,
\nonumber \\ 
    \ave{\,\varphi^\dagger_{AA'}(x)} &= \Delta^*_A(x)\,\delta_{AA'}\,,
\end{alignat}
where the ``gap function'' $\Delta_A(x)$ is now a classical field. 
Here we assume that the condensation takes place only in the 
diagonal flavor-color components of the gap matrix, $A=A'$, 
as in the standard ansatz for the CFL or the 2SC phase.
Note, however, that we retain the full space-time dependence of the
field. 

Introducing Nambu-Gor'kov bispinors,
\beq
\Psi(x)
=
\frac{1}{\sqrt{2}}
\left(
\begin{array}{c}
q(x)\\
q_C(x)
\end{array}
\right)
\,,\quad
\bar\Psi(x)
=
\frac{1}{\sqrt{2}}
\left(
\bar q(x), 
\bar q_C(x)
\right)\,,
\eeq
we obtain the effective mean-field Lagrangian
\beq
    \mathcal{L}_\mathit{MF}(x) = \bar\Psi(x)\,S^{-1}(x)\,\Psi(x) 
    \,-\, \frac{ 1}{4H} \sum_A |\Delta_A(x)|^2\,,
\label{LMF}
\eeq
with the inverse dressed quark propagator
\beq
     S^{-1}(x) = 
\left(
\begin{array}{cc}
i\dslash+\muslash & \hat\Delta(x)\, \gamma_{5}\\
-\hat\Delta^{*}(x)\, \gamma_{5} & i\dslash-\muslash
\end{array}
\right)\,.
\eeq
Here we used the more compact notation
\beq
    \hat\Delta(x) = \sum_{A} \Delta_{A}(x) \,\tau_{A}\lambda_{A}\,,
\label{Deltax}
\eeq
i.e., $\hat\Delta(x)$ is a matrix in color and flavor space.

\subsection{Thermodynamic potential}
\label{sec:Omega}

We now consider a static crystalline structure with a unit cell spanned by
three linearly independent vectors $\vec a_1$, $\vec a_2$, and  $\vec a_3$.
This means, the gap matrix $\hat\Delta(x)$ is time independent and periodic in 
space,
\beq
     \hat\Delta(x) \equiv \hat\Delta(\vec x) = \hat\Delta(\vec x + \vec a_i)\,,
     \quad i = 1,2,3\,.
\eeq
Hence, $\hat\Delta$ can be decomposed into a discrete set of Fourier 
components,
\beq
    \hat\Delta(x) = \sum_{q_k} \hat\Delta_{q_k}\,e^{-iq_k\cdot x}\,,
\label{Deltaq}
\eeq
where the allowed momenta are given by the conditions
\beq
    q_k = \left(\begin{array}{c} 0 \\ \vec q_k \end{array} \right)\,,
    \qquad \vec q_k\cdot\vec a_i = 2\pi\,N_{ki}\,,
\label{qk}
\eeq
for $N_{ki} \in \Z$.
These momenta form a reciprocal lattice ($R.L.$) in momentum space.

For the bispinors $\Psi$ and $\bar\Psi$ we consider a finite quantization
volume $V$ with periodic boundary conditions.
Working at finite temperature $T$ and employing Matsubara formalism
the (imaginary) time variable is restricted to a finite interval as well,
$0 \leq \tau = it \leq 1/T$. Hence, the allowed energies and three-momenta
both are discrete and we have the Fourier decompositions
\begin{alignat}{1}
  \Psi(x) &= \frac{1}{\sqrt{V}} \sum_{p_n} \Psi_{p_n}\, e^{-ip_n\cdot x}\,,
\nonumber\\
  \bar\Psi(x) &= \frac{1}{\sqrt{V}} \sum_{p_n} \bar\Psi_{p_n}\, e^{ip_n\cdot x}\,.
\label{Psip}
\end{alignat}
Here we have explicitly taken out a normalization factor $1/\sqrt{V}$ to
have dimensionless Fourier components $\Psi_{p_n}$ and $\bar\Psi_{p_n}$.

For a consistent description of the crystal, the quantization volume
should contain an integer number of unit cells. Without loss of generality
we therefore assume that $V$ is spanned by the vectors $N\,\vec a$, where $N$ 
is a positive integer. Then the allowed momenta are given by
\beq
    p_n = \left(\begin{array}{c} i\omega_{p_n} \\ \vec p_n \end{array} 
   \right)\,,
    \qquad \vec p_n\cdot\vec a_i = 2\pi\,\frac{N_{ni}}{N}\,,
\label{pn}
\eeq
with $\omega_{p_n}$ being fermionic Matsubara frequencies and 
$N_{ni} \in \Z$.
Comparing this with \eq{qk}, we see that the three-momenta form a mesh
which, in each direction, is $N$ times finer than the reciprocal lattice
of the crystal. 
Later we will take the infinite volume limit, $N\rightarrow\infty$,
where the set of allowed three-momenta becomes continuous.

The thermodynamic potential per volume is given by
\beq
\Omega(T,\mu) =
-\frac{T}{V}\ln {\mathcal Z}(T,\mu)\,,
\eeq
where
\beq
    {\mathcal Z}(T,\mu) =
    \int\mathcal{D}\Psi\,\mathcal{D}\bar{\Psi} \;e^{\mathcal{S}}
\eeq
is the grand canonical partition function with the Euclidean action
\beq
    \mathcal{S} = \int_0^{1/T} \hspace{-2mm}d\tau \int_V d^3x\; 
    \mathcal{L}(x^0=-i\tau,\vec x)\,. 
\eeq 

Inserting the Fourier decompositions, 
\eqs{Deltaq} and (\ref{Psip}), into \eq{LMF} and turning out the 
integrals, we obtain in mean-field approximation
\beq
    \mathcal{S}_\mathit{MF} = 
    \frac{1}{T} \sum_{p_m,p_n} \bar\Psi_{p_m} S^{-1}_{p_m,p_n} \Psi_{p_n}
    \,-\, \frac{1}{4H} \frac{V}{T} \sum_A\sum_{q_k} |\Delta_{A,q_k}|^2\,,
\label{SMF}
\eeq
where
\begin{alignat}{1}
 &S^{-1}_{p_m,p_n} = 
\nonumber\\
 &\left(
\begin{array}{cc}
(\pslash_n+\muslash)\,\delta_{p_m,p_n} & 
\sum_{q_k}\hat\Delta_{q_k}\gamma_{5}\,\delta_{q_k,p_m-p_n}\\
-\sum_{q_k}\hat\Delta^*_{q_k}\gamma_{5}\,\delta_{q_k,p_n-p_m}& 
(\pslash_n-\muslash)\,\delta_{p_m,p_n}
\end{array}
\right)
\label{Sinv}
\end{alignat}
is the $(p_m,p_n)$-component of the inverse quark propagator in
momentum representation.
Note that in general $S^{-1}$ is not diagonal in momentum space
because the condensates $\hat\Delta_{q_k}$ couple different momenta.
Physically, this corresponds to processes like the absorption of 
a quark with momentum $p_n$ by the condensate together with the 
emission of an antiquark or a hole with momentum $p_m = p_n + q_k$.
This is only possible because the inhomogeneous diquark condensates
carry momentum. In the homogeneous case, $\hat\Delta(x) = \mathit{const.}$,
only the momentum component $q_k = 0$ exists, and the in- and outgoing
quark momenta are equal. 
While this is no longer true for our inhomogeneous ansatz, the fact 
that we consider a static solution still guarantees that the {\it energy}
of the quark is conserved, see \eq{qk}. This means, $S^{-1}$ is still 
diagonal in the Matsubara frequencies $\omega_{p_n}$.

Since the action, \eq{SMF}, is bilinear in the fields (plus a field 
independent term) the mean field thermodynamic potential is readily
evaluated. We obtain
\beq
    \Omega_{MF}(T,\mu) \,=\, \Omega_{0}(T,\mu)     
    \,+\, \frac{1}{4H} \sum_A\sum_{q_k} |\Delta_{A,q_k}|^2\,,
\label{eq:menfieldOm}
\eeq
with
\beq
    \Omega_{0}(T,\mu) \,=\, -\frac{1}{2}\frac{T}{V} 
    \mathrm{Tr}\ln\,\frac{S^{-1}}{T}\,,
\label{Om0}
\eeq
where the trace is to be taken over the Nambu-Gor'kov,
Dirac, color, flavor, and momentum components of the inverse propagator.
The factor $\frac{1}{2}$ in front corrects for overcounting due to the
artificial doubling of the degrees of freedom in
Nambu-Gor'kov formalism. 

As pointed out above, the inverse propagator is diagonal in the energy
components. This allows us to perform the energy trace, i.e., the 
Matsubara sum in the usual way. To that end we write
\beq
    S^{-1}_{p_m,p_n} = \gamma^0\,(i\omega_{p_n} - 
    {\cal H}_{\vec p_m,\vec p_n})\, \delta_{\omega_{p_m},\omega_{p_n}}\,,
\label{SH}
\eeq
with the effective Hamilton operator
\begin{alignat}{1}
 &{\cal H}_{\vec p_m,\vec p_n} = 
\nonumber\\[1mm]
 &\left(
\begin{array}{cc}
(\gamma^0\pvslash_n-\mu)\,\delta_{\vec p_m,\vec p_n} & 
-\sum_{\vec q_k}\hat\Delta_{q_k}\gamma^0 \gamma_{5}\,\delta_{\vec q_k,\vec p_m-\vec p_n}
\\
\sum_{\vec q_k}\hat\Delta^*_{q_k}\gamma_0\gamma_{5}\,\delta_{\vec q_k,\vec p_n-\vec p_m}
& 
(\gamma^0\pvslash_n+\mu)\,\delta_{\vec p_m,\vec p_n}
\end{array}
\right)\,,
\label{Hmn}
\end{alignat}
which does not depend on $\omega_{p_n}$. 
Here we have introduced the notation $\pvslash = \vec\gamma\cdot\vec p$.

Since ${\cal H}$ is hermitian, it can
in principle be diagonalized. We can then employ the formula
\beq
T\sum_{\omega_{p_n}}\ln\left(\frac{i\omega_{p_n}+E_\lambda}{T}\right)
=
\frac{E_\lambda}{2} + 
T \,\ln \left(1 + e^{-E_\lambda/T} \right)
\eeq
to turn out the Matsubara sum. In this way we obtain
\beq
    \Omega_{0}(T,\mu) \,=\, -\frac{1}{4V} \sum_{\lambda}
    \left( E_\lambda + 2T\,\ln \left(1 + 2e^{-E_\lambda/T} \right) \right)\,,
\label{Om0Mat}
\eeq
where the sum is over all eigenvalues $E_\lambda$
of ${\cal H}$ in Nambu-Gor'kov, Dirac, color, flavor, and three-momentum 
space.

\eq{Om0Mat} is formally the same as for homogeneous condensates. 
In practice, since ${\cal H}$ is not diagonal in three-momentum space, 
its diagonalization is of course much more difficult in the 
inhomogeneous case. However, as a consequence of the periodicity of the 
crystal, ${\cal H}$ can be brought into block diagonal form.
As obvious from \eq{Hmn}, only those quark momenta $\vec p_m$ and $\vec p_n$ 
are coupled which differ by a momentum $\vec q_n$ belonging to the
$R.L.$ of the crystal. On the other hand we have seen earlier that,
for a quantization volume $V$ containing $N^3$ unit cells,
the mesh of allowed quark momenta is $N^3$ times finer than the 
$R.L.$, cf.~\eqs{qk} and (\ref{pn}).
Therefore, ${\cal H}$ can be decomposed into $N^3$ independent blocks in 
momentum space. The sum over the eigenstates $\lambda$ in \eq{Om0Mat} thus 
separates into a sum over the different blocks times
a sum over the eigenstates of each block.
The reader may recognize that this structure is deeply related to the Bloch 
theorem which basically says that eigenfunctions can be labelled by a vector 
in the Brillouin zone  ($B.Z.$) and that eigenfunctions for different vectors 
are orthogonal.

More precisely, we write
\beq
    \vec p_m = \vec k_m + \vec q_m\,, \qquad
    \vec p_n = \vec k_n + \vec q_n\,, 
\label{pmpn}
\eeq
where $\vec q_m$ and $\vec q_n$ are elements the $R.L.$ 
and $\vec k_m$ and $\vec k_n$ belong to the $B.Z.$
Then $\vec p_m$ and $\vec p_n$ are coupled only if $\vec k_m =\vec k_n$.
For each vector $\vec k_n$ in the $B.Z.$ we therefore define a projector
\beq
\left(P_{\vec{k}_{n}}\right)_{\vec{p}_{m},\vec{p}_{n}}
=
\sum_{\vec{q}_{m},\vec{q}_{n}\in R.L.}
\delta_{\vec{p}_{m}-\vec{k}_{n},\vec{q}_{m}}\,
\delta_{\vec{p}_{n}-\vec{k}_{n},\vec{q}_{n}}\,,
\eeq
which commutes with ${\cal H}$ and which projects out the block of coupled
momenta related to $\vec k_n$. 
The effective Hamilton operator ${\cal H}$, \eq{Hmn}, can thus be written
as a direct sum 
\beq
    {\cal H} = \sum_{\vec k_n \in B.Z.}  {\cal H}(\vec k_n)~,
\label{Hk}
\eeq
where
\beq
\big({\cal H}(\vec k_n)\big)_{\vec p_m,\vec p_n} = 
 \left(
\begin{array}{cc}
(\gamma^0\pvslash_n-\mu)\,\delta_{\vec p_m,\vec p_n} & 
-\hat\Delta_{p_m-p_n}\gamma^0 \gamma_{5}
\\
\hat\Delta^*_{p_n-p_m}\gamma_0\gamma_{5}
& 
(\gamma^0\pvslash_n+\mu)\,\delta_{\vec p_m,\vec p_n}
\end{array}
\right)
\label{Hkmn}
\eeq
is the non-trivial part of $P_ {\vec k_n}{\cal H}$.
Here $\vec p_m$ and $\vec p_n$ are restricted to the corresponding
subspace.

Accordingly, we obtain for the thermodynamic potential
\begin{alignat}{1}
    &\Omega_{0}(T,\mu) \;= 
\nonumber\\
    &-\frac{1}{4V} \sum_{\vec k_n\in B.Z.} 
     \sum_{\lambda}
    \left( E_\lambda(\vec k_n)
    + 2T\,\ln \left(1 + e^{-E_\lambda(\vec k_n)/T} \right) \right)\,,
\end{alignat}
where $E_{\lambda}(\vec{k}_{n})$ are the non-trivial eigenvalues of
${\cal H}(\vec k_n)$.

Finally, we can take the infinite volume limit,
\beq
\frac{1}{V}\sum_{\vec{k}_{n}\in B.Z.}
\rightarrow
\int\limits_{B.Z.}\!\!\frac{d^3k}{(2 \pi)^3}
\,.
\eeq
We then obtain
\begin{alignat}{1}
&\Omega_{0}(T,\mu) \;=
\nonumber\\
&-\frac{1}{4}
\int\limits_{B.Z.}\!\!\frac{d^3k}{(2 \pi)^3}
\sum_{\lambda}
\left\{
E_\lambda(\vec k)
+2T \ln\, \left(1+e^{-E_\lambda(\vec k)/T}\right)
\right\}
\,.
\label{Om0cont}
\end{alignat}
In particular for $T=0$, we have  
\beq
\Omega_{0}(T=0,\mu)
=
-\frac{1}{4}
\int\limits_{B.Z.}\!\!\frac{d^3k}{(2 \pi)^3}
\sum_{\lambda}
\left\vert E_{\lambda}(\vec{k})\right\vert
\,.
\eeq
These formulas are of course consistent with the homogeneous
case. In this limit the ``reciprocal lattice'' only consists of the point 
$\vec q = 0$ and the $B.Z.$ is the entire three-momentum space.

Before closing this section, let us give an interpretation of the
momenta in the equations above. Inspecting the upper right
Nambu-Gor'kov component in \eq{Hkmn}, we see that $\hat\Delta_{p_m-p_n}$
couples an incoming hole with momentum $\vec p_n$ to an outgoing 
particle with momentum $\vec p_m$. This means, the condensate contains
a fermion pair with momenta $-\vec p_n$ and $\vec p_m$, respectively. 
Writing $\vec p_m = \vec k + \vec q_m$ and 
$-\vec p_n = -\vec k - \vec q_n$ with $\vec k \in B.Z.$ and 
$\vec q_m, \vec q_n \in R.L.$, 
we see that $\vec p_m - \vec p_n$ is just the total momentum of the 
pair, whereas $2\vec k$ is the relative momentum modulo momenta
of the $R.L.$

\subsection{Regularization}
\label{sec:regularization}

The above expressions for the thermodynamic potential are quartically
divergent if the integral and the sum
are left unconstrained. Therefore, we have to specify a regularization
procedure to get a well defined result. 
Since later we want to compare the free energies of inhomogeneous
and homogeneous solutions, it is of course crucial to regularize both
cases in a consistent way. 

We do not want to attach a physical meaning to the regularization
scheme. Instead we think of a local theory, which -- at a given order
in some power-counting scheme -- can be ``renormalized'' 
by adding a finite number of local operators. 
The corresponding counter terms should be determinable in the homogeneous 
phase and be expressible through physical observables.
As a consequence, the Ginzburg-Landau coefficients, which can be derived
from the mean-field thermodynamic potential, 
should depend on the regularization only indirectly through physical 
observables, like the BCS gap at a given chemical potential.


Given these constraints, it is not obvious how to generalize a three-momentum cutoff regularization to inhomogeneous phases and as discussed in Appendix~\ref{app:GL} the most naive approach of restricting the momenta $\vec{k}$ of external and internal quarks  to $\vert\vec{k}\vert<\Lambda$ or $k_F - \Lambda \leq |\vec k| \leq k_F + \Lambda$ does not meet our requirements.

We therefore suggest a Pauli-Villars-like regularization scheme, which we introduce via a proper-time regularization of the functional logarithm in the thermodynamic potential (see Eq.~\ref{Om0}) and which does therefore not rely on homogeneous ground states.

In a first step we go back to \eqs{Om0} and (\ref{SH}) and combine
positive and negative Matsubara frequencies to get
\begin{alignat}{1}
    T\sum_n \ln\,(i\omega_n - \mathcal{H})
    &=
    \frac{T}{2}\sum_n \ln(i\omega_n - \mathcal{H})(-i\omega_n - \mathcal{H})
\nonumber\\
    &= \frac{T}{2}\sum_n \ln\,A~,
\end{alignat}
where $A = \omega_n^2 + \mathcal{H}^2$ is a hermitian operator with
positive eigenvalues. 
Next we replace the logarithm by its Schwinger proper-time representation,
\beq
    \ln\,A \rightarrow -\int_0^\infty \frac{d\tau}{\tau} f(\tau) e^{-\tau A}~,
\label{SPT}
\eeq
where we introduced a blocking function $f(\tau)$ as a regulator. 
Thus, our regularization scheme is defined by specifying $f(\tau)$.

The most simple prescription would be to put a lower bound in the 
proper-time variable, $f(\tau) = \theta(\tau - 1/\Lambda^2)$.
However, as we would like to keep a structure which allows us to
perform the Matsubara sum analytically, we prefer the function
\beq
    f(\tau) = 1 - 2 e^{-\tau\Lambda^2} + e^{-2\tau\Lambda^2}~.
\label{ftau}
\eeq
Inserting this into \eq{SPT}, this amounts to the replacement
\beq
    \ln\,A \rightarrow \ln\,A -2\ln\,(A+\Lambda^2) + \ln\,(A+2\Lambda^2)~,
\label{PTPV}
\eeq
and we can carry out the Matsubara sum as before. 
Then the regularized version of \eq{Om0cont} reads:
\begin{alignat}{1}
\Omega_{0}(T,\mu) &=
-\frac{1}{4}
\int\limits_{B.Z.}\!\!\frac{d^3k}{(2 \pi)^3}
\sum_{\lambda}
\nonumber\\
&\times \sum_{j=0}^2 c_j
\left\{
E_{\lambda,j}(\vec k)
+2T \ln\, \left(1+e^{-E_{\lambda,j}(\vec k)/T}\right)
\right\}
\,,
\label{Om0reg}
\end{alignat}
where $c_0 = c_2 = 1$, $c_1 = -2$, and
\beq
    E_{\lambda,j}(\vec k) = 
    \sqrt{E_\lambda^2(\vec k) + j \Lambda^2}~.
\label{Ereg}
\eeq
\eqs{PTPV} and (\ref{Om0reg}) are reminiscent of Pauli-Villars 
regularization. 
Note, however, that according to \eq{Ereg} we replace the
{\it free energies} in a Pauli-Villars-like manner, 
which is not exactly the same as introducing regulator particles 
with large masses, as in the standard Pauli-Villars regularization. 

The two regulator terms generated by the blocking function \eq{ftau}
regularize quadratic divergencies. This is sufficient to get finite
results for derivatives of the thermodynamic potential, like the quark
number density $-\frac{\partial\Omega}{\partial\mu}$ or the derivatives
$\frac{\partial\Omega}{\partial\hat\Delta_{q_k}}$, which appear in the
gap equations.
On the other hand, since the unregularized thermodynamic potential is
quartically divergent, it remains logarithmically divergent, even
after the regularization. It should be kept in mind, however, that
the thermodynamic potential is physically meaningful only up to a 
constant. We can therefore subtract the remaining divergency by 
calculating the difference to some reference point, like 
the ground state in vacuum or simply the normal conducting phase
at some given chemical potential.

\subsection{Simplified model}
\label{sec:simple}

The general framework for the mean-field thermodynamic potential 
derived in Sec.~\ref{sec:Omega}, together with the regularization 
procedure suggested in Sec.~\ref{sec:regularization}, is one of our
central results.
It may serve as a starting point for extensive studies of the
phase diagram of strongly interacting matter in future investigations.
In most of the remainder of the present article, we will illustrate the
power of our approach by numerical examples in a simplified version
of our model.

\subsubsection{Dirac structure}

In a first step, we derive an approximation to our model, which should be 
valid at high densities.
Starting point is the effective Hamilton operator ${\cal H}$, \eq{Hmn}.
As we have discussed, ${\cal H}$ is block diagonal in 
momentum space, and we can in fact concentrate on a single block 
${\cal H}(\vec k)$, related to the offset momentum $\vec k \in B.Z.$, 
cf.~\eq{Hkmn}. However, for the sake of notational brevity, we will drop 
the argument $(\vec k)$ in the following.

In order to reduce the complexity, and guided by high-density effective
theory, we want to get rid of the Dirac structure in ${\cal H}$. 
We first note that the eigenvalue spectrum does not
change under unitary transformations. Choosing
\beq
    U = \left( 
        \begin{array}{cc} 1 & 0 \\ 0 & \gamma^0\gamma_5 \end{array}
        \right) = -U^\dagger = -U^{-1}~,
\eeq
we obtain ${\cal H}' = U^\dagger \, {\cal H} \, U$ with
\beq
 {\cal H}'_{\vec p_m,\vec p_n} = \left(
\begin{array}{cc}
(\gamma^0\pvslash_n-\mu)\,\delta_{\vec p_m,\vec p_n} &  \hat\Delta_{p_m-p_n}
\\
\hat\Delta^*_{p_n-p_m} & -(\gamma^0\pvslash_n+\mu)\,\delta_{\vec p_m,\vec p_n}
\end{array}
\right). 
\eeq
In homogeneous phases, a standard method to diagonalize the remaining Dirac
structure is to employ energy projectors, 
\beq
    \Lambda_{\hat p}^{\pm} = \frac{1}{2}(1 \pm \gamma^0 \phslash)~,
\eeq
where $\hat p = {\vec p}/{p}$ with $p = |\vec p|$.
We can then reexpress
\beq
    \gamma^0\pvslash_n-\mu = (p_n - \mu) \Lambda_{\hat p}^{+} -
    (p_n + \mu) \Lambda_{\hat p}^{-}~,
\eeq
and ${\cal H}'$ can be decomposed into a positive and
a negative energy part, which act on orthogonal subspaces of the Hilbert 
space. One can therefore find a new basis, where these positive and 
negative energy parts decouple. 

However, as the $\Lambda_{\hat p}^{\pm}$ are only projectors for states
with the same momentum direction $\hat p$, the method described
above, does not work exactly in inhomogeneous phases, where different 
momenta are coupled by the gap functions.
In fact, we can still remove the Dirac structure from the diagonal 
momentum components in this way, but at the same time new Dirac 
structures appear in the off-diagonal components. 

We can work this out explicitly by performing a unitary transformation
${\cal H}'' = V^\dagger\,{\cal H}'\,V$
which diagonalizes the Dirac part of the diagonal momentum components
${\cal H}'_{\vec p_m,\vec p_m}$. 
The result reads
\beq
     {\cal H}'' = {\cal H}_{1}'' + {\cal H}_{2}''~,
\eeq
where 
\begin{widetext}
\beq
({\cal H}''_{1})_{\vec p_m,\vec p_n} = 
\left(
\begin{array}{cccc}
(p_m-\mu)\,\delta_{\vec p_m,\vec p_n} &  \hat\Delta_{p_m-p_n} &  &
\\
\hat\Delta^*_{p_n-p_m} & -(p_m-\mu)\,\delta_{\vec p_m,\vec p_n} &  &
\\
& & -(p_m+\mu)\,\delta_{\vec p_m,\vec p_n} &  \hat\Delta_{p_m-p_n}
\\
& & \hat\Delta^*_{p_n-p_m} & (p_m+\mu)\,\delta_{\vec p_m,\vec p_n}
\end{array}
\right)
\otimes \unity_2
\eeq
\end{widetext}
is diagonal in Dirac space. 
Here we have reordered the lines and columns to make the diagonal
structure obvious.
The non-diagonal $2 \times 2$ blocks now describe the Nambu-Gor'kov
structure. 
The $2 \times 2$ identity matrix on the right is 
related to the spin degeneracy of the problem and is, thus,
part of the Dirac structure.

In this basis, the remaining part of the transformed Hamiltonian is given by
\begin{widetext}
\beq
({\cal H}''_{2})_{\vec p_m,\vec p_n} = 
\left(
\begin{array}{cccc}
0 &  \hat\Delta_{p_m-p_n} & 0 & -\hat\Delta_{p_m-p_n}
\\[1mm]
\hat\Delta^*_{p_n-p_m} & 0 & -\hat\Delta^*_{p_n-p_m} & 0
\\[1mm]
0 &  -\hat\Delta_{p_m-p_n} & 0 & \hat\Delta_{p_m-p_n}
\\[1mm]
-\hat\Delta^*_{p_n-p_m} & 0 & \hat\Delta^*_{p_n-p_m} & 0
\end{array}
\right)
\otimes  
\frac{1}{2} \Big( (\hat p_m \cdot \hat p_n - 1)\,\unity_2  
+ i (\hat p_m \times \hat p_n) \cdot \vec\sigma \Big)~,
\eeq
\end{widetext}
where $\vec\sigma$ denotes the Pauli matrices, indicating a non-trivial
spin structure. Obviously, ${\cal H}''_{2}$ is not diagonal in Dirac space.
However, the components of ${\cal H}''_{2}$ vanish for parallel momenta, 
$\hat p_m = \hat p_n$.
This reflects the fact that states with parallel momenta have the same
energy projectors $\Lambda_{\hat p}^\pm$.

In the following, we will neglect ${\cal H}''_2$. We expect that this is
a good approximation at very high densities where the physics is dominated 
by collinear scattering near the Fermi surface,
$|\vec p_m - \vec p_n| \ll p_m \sim p_n \sim \mu$. 
Neglecting the antiparticle contributions as well, we obtain
\begin{alignat}{1}
&\Omega_{0}(T,\mu)
\nonumber\\
&=
-\frac{1}{2}
\int\limits_{B.Z.}\!\!\frac{d^3k}{(2 \pi)^3}
\sum_{\lambda}
\left\{
E_\lambda(\vec k)
+2T \ln\, \left(1+e^{-E_\lambda(\vec k)/T}\right)
\right\}
\nonumber\\ & + \quad \mathit{regulator\; terms}\,,
\end{alignat}
where $E_\lambda(\vec k)$ are now the eigenvalues of the 
``high-density effective Hamiltonian''
\beq
({\cal H}_\mathit{HDE})_{\vec p_m,\vec p_n} = 
\left(
\begin{array}{cc}
(p_m-\mu)\,\delta_{\vec p_m,\vec p_n} &  \hat\Delta_{p_m-p_n}
\\
\hat\Delta^*_{p_n-p_m} & -(p_m-\mu)\,\delta_{\vec p_m,\vec p_n}
\end{array}
\right)
\eeq
for a given offset momentum $\vec k$.

\subsubsection{Color-flavor structure}

The second simplification concerns the color-flavor structure of the model.
In Sec.~\ref{Lagrangian} we have chosen a form of the gap matrix which is 
capable to describe the most common phases in color superconductivity,
see \eq{Deltax}. For instance, the CFL phase corresponds to the case
$\Delta_2 = \Delta_5 = \Delta_7 \equiv \Delta$.

In the following, we restrict ourselves to the 2SC pairing pattern, 
which is defined by $\Delta_2 = \Delta$ and $\Delta_5 = \Delta_7 = 0$.
Moreover, since we are only interested in the effect of the pairing 
relative to the normal conducting phase, we can omit those colors
and flavors which do not participate in the pairing (i.e., 
strange quarks and, using standard nomenclature, blue quarks).
Thus, the remaining Hamiltonian is a $4 \times 4$ matrix in color-flavor 
space which can trivially be decomposed into four separate blocks.

We assume that the chemical potential may be different for up and
down quarks, 
\beq
    \mu_u = \bar\mu + \delta\mu~, \quad \mu_d = \bar\mu - \delta\mu~,
\label{deltamu}
\eeq
but does not depend on color.
We then obtain
\beq
     {\cal H}_\mathit{HDE} = {\cal H}_{\Delta,\delta\mu} \oplus 
     {\cal H}_{-\Delta,\delta\mu} \oplus{\cal H}_{\Delta,-\delta\mu} 
    \oplus{\cal H}_{-\Delta,-\delta\mu}~,
\label{HHDE}
\eeq
where
\begin{alignat}{1}
&({\cal H}_{\Delta,\delta\mu})_{\vec p_m,\vec p_n} = 
\nonumber\\[1mm]
&\left(
\begin{array}{cc}
(p_m-\bar\mu-\delta\mu)\,\delta_{\vec p_m,\vec p_n} &  \Delta_{p_m-p_n}
\\
\Delta^*_{p_n-p_m} & -(p_m-\bar\mu+\delta\mu)\,\delta_{\vec p_m,\vec p_n}
\end{array}
\right)~.
\label{HDm}
\end{alignat}
Of course, the eigenvalues of ${\cal H}_{\Delta,\delta\mu}$ do not
depend on the overall sign of $\Delta$. 
Moreover, at least in all cases to be considered in this article,
replacing $\delta\mu$ by $-\delta\mu$ amounts to a replacement of the
eigenvalue spectrum $\{E_\lambda(\vec k)\}$ by $\{-E_\lambda(\vec k)\}$
(see Appendix~\ref{app:symevs}).
Hence, since the thermodynamic potential depends only on the moduli of the
eigenvalues, each of the four blocks contributes equally, and we
only need to determine the eigenvalues of ${\cal H}_{\Delta,\delta\mu}$.

\subsubsection{Gap equations}

Summarizing our main equations, including the approximations
introduced above, the regularized mean-field thermodynamic potential is 
given by
\beq
    \Omega(T, \bar\mu, \delta\mu) \,=\, \Omega_{0}(T, \bar\mu, \delta\mu)   
    \,+\, \frac{1}{4H} \sum_{q_k} |\Delta_{q_k}|^2\,,
\label{Omdmu}
\eeq
with 
\begin{alignat}{1}
&\Omega_{0}(T, \bar\mu, \delta\mu)
= 
-2 \int\limits_{B.Z.}\!\!\frac{d^3k}{(2 \pi)^3}
\sum_{\lambda}
\nonumber\\
&\hspace{10mm}\times \sum_{j=0}^2 c_j
\left\{
E_{\lambda,j}(\vec k)
+2T \ln\, \left(1+e^{-E_{\lambda,j}(\vec k)/T}\right)
\right\}
\,,
\label{Om0dmu}
\end{alignat}
where $c_0 = c_2 = 1$, $c_1 = -2$, and
$E_{\lambda,j}(\vec k) = \sqrt{E_\lambda^2(\vec k) + j \Lambda^2}$
as in \eq{Ereg}. $E_{\lambda}(\vec k)$ are now the eigenvalues of 
${\cal H}_{\Delta,\delta\mu}(\vec k)$, \eq{HDm}.
At $T=0$, which will be our main focus, this becomes
\begin{alignat}{1}
&\Omega_{0}(T=0, \bar\mu, \delta\mu)
= 
-2 \int\limits_{B.Z.}\!\!\frac{d^3k}{(2 \pi)^3}
\sum_{\lambda}\sum_{j=0}^2 c_j
\left|E_{\lambda,j}(\vec k)\right|
\,.
\label{Om0dmuT0}
\end{alignat}
As discussed earlier, in order to get a finite result for $\Omega$,
one still has to subtract an infinite constant. In this paper, we
will always consider the free-energy difference to the normal phase,
i.e., the phase with $\Delta \equiv 0$ at the same value of 
$T$, $\bar\mu$, and $\delta\mu$. This quantity is finite.

Finally, before coming to the numerical results, we want to discuss
the gap equations. 
As in the homogeneous case we have to minimize the thermodynamic 
potential, which means that we have to solve the equations
\beq
\label{eq:gap1}
0
=
\frac{\partial\Omega}{\partial \Delta_{q_k}^{*}}
=
\frac{\partial\Omega_0}{\partial \Delta_{q_k}^{*}}
+
\frac{\Delta_{q_k}}{4H}
\eeq
for all Fourier components. From \eq{Om0dmu}, we obtain
\begin{alignat}{1}
\frac{\partial\Omega_0}{\partial \Delta_{q_k}^{*}}
= 
-2 \int\limits_{B.Z.}\!\!\frac{d^3k}{(2 \pi)^3}
&\sum_{\lambda}
\frac{\partial E_{\lambda}}{\partial \Delta_{q_k}^{*}}
\nonumber\\
\times \quad&\sum_{j=0}^2 c_j
\Big(1 -2n(E_{\lambda,j})\Big)\,
\frac{E_{\lambda,j}}{E_{\lambda}}\,,
\label{dOm0dDelta}
\end{alignat}
where $n(E) = [\exp(E/T) + 1]^{-1}$ is a Fermi function.
To evaluate this further, we use that $E_{\lambda}$ are the
eigenvalues of ${\cal H}_{\Delta,\delta\mu}$.
This means, there are unitary matrices $U$ so that
\beq
    \tilde{\mathcal{H}}_{\Delta,\delta\mu}
    =
    U^{-1}\,\mathcal{H}_{\Delta,\delta\mu}\,U
\eeq
is a diagonal matrix with 
$(\tilde{\mathcal{H}}_{\Delta,\delta\mu})_{\lambda\lambda} = E_\lambda$.
Hence,
\beq
    \frac{\partial E_{\lambda}}{\partial \Delta_{q_k}^{*}}    
    =
    \frac{\partial}{\partial \Delta_{q_k}^{*}}\,
    \left(U^{-1}\,\mathcal{H}_{\Delta,\delta\mu}\,U\right)_{\lambda\lambda}
    =
    \left(U^{-1}\,P_{q_{k}}\,U\right)_{\lambda\lambda}\,,
\eeq
where $P_{q_{k}}=\frac{\partial}{\partial\Delta_{q_{k}}^{*}}
\mathcal{H}_{\Delta,\delta\mu}$ is a known $\Delta$- and 
$\vec k$-independent matrix (see \eq{Pqk} below).
Note that the terms related to the derivatives of $U^{-1}$ and $U$ 
cancel each other.
Writing the matrix $U=(w_{1},\dots)$ in terms of the 
eigenvectors $w_\lambda$ to the eigenvalues $E_\lambda$,
this yields
\beq
    \frac{\partial E_{\lambda}}{\partial \Delta_{q_k}^{*}}    
    = w_{\lambda}^\dagger P_{q_{k}} w_{\lambda}\,.
\eeq
In fact, this formula is nothing but first-order perturbation theory
for the modification of the eigenvalues under a small perturbation 
$\delta\mathcal{H}_{\Delta,\delta\mu} = P_{q_{k}}\delta \Delta_{q_k}$.
If we insert this into \eq{dOm0dDelta}, we can then
rewrite \eq{eq:gap1} into the gap equations
\begin{alignat}{1}
\Delta_{q_k}
= 
8H \int\limits_{B.Z.}\!\!\frac{d^3k}{(2 \pi)^3}
&\sum_{\lambda}
w_{\lambda}^\dagger P_{q_{k}} w_{\lambda}
\nonumber\\
\times \quad&\sum_{j=0}^2 c_j
\Big(1 -2n(E_{\lambda,j})\Big)\,
\frac{E_{\lambda,j}}{E_{\lambda}}\,.
\label{eq:gap2}
\end{alignat}
These equations are the basis for our numerical analysis.

At this point, we would like to note that 
the matrix $P_{q_{k}}$ connects momenta differing by $\vec{q}_k$.
In fact, from \eq{HDm} we get
\beq (P_{q_{k}})_{\vec p_m,\vec p_n} = 
\left(\begin{array}{cc} 0 & 0 \\ 1 & 0 \end{array}\right)\,
\delta_{\vec p_n-\vec p_m,\vec q_{k}}
\label{Pqk}
\,.
\eeq
Thus, denoting the Nambu-Gor'kov components of the eigenvectors
as $w_\lambda =(u_\lambda,v_\lambda)$ 
and indicating the momentum components explicitly, the gap equations
read
\beq
\Delta_{q_k}
= 
8H \int\limits_{B.Z.}\!\!\frac{d^3k}{(2 \pi)^3}
\sum_{\lambda, \vec p_n}\,
(v_{\lambda}^\dagger)_{\vec p_n - \vec q_k}\,(u_{\lambda})_{\vec p_n}
\,\Big(1 -2n(E_{\lambda})\Big)\,,
\eeq
where we have omitted the regulator terms for simplicity.
Fourier transforming this convolution with the conventions given  
in Eqs.~(\ref{Deltaq},\ref{Psip}) we get
\beq
\Delta(x)
= 
8HV \int\limits_{B.Z.}\!\!\frac{d^3k}{(2 \pi)^3}
\sum_{\lambda}\,
v_{\lambda}^\dagger(x)\,u_{\lambda}(x)
\,\Big(1 -2n(E_{\lambda})\Big)\,,
\eeq
where $V$ is the volume of the unit cell.
This is precisely the selfconsistency condition for the Bogoliubov-de Gennes 
equation~\cite{deGennes66} when again exploiting Bloch's theorem.
It is obvious that this relation is not restricted to the simplified
model we discussed here, but an extension to the 
general case is straight forward.
Note however that our regularization procedure and therefore also our 
numerical calculations are tied in momentum space.


\section{Numerical results}
\label{sec:results}

In this section we want to discuss numerical
calculations performed within the simplified model of Sec.~\ref{sec:simple}.
We restrict ourselves to $T=0$ and to a fixed average chemical
potential $\bar\mu = 400$~MeV.
Then $\delta\mu$ is the only remaining external variable and we will
drop the arguments $T$ and $\bar\mu$ in the following.

Our model has two parameters, namely the coupling constant $H$ and the 
cutoff parameter $\Lambda$.
We remind that $\Lambda$ restricts the free energies and not the momenta. 
Thus, the most relevant excitations around the Fermi surface are always
included, and there is no need for $\Lambda$ to be larger than the
chemical potential. Of course, $\Lambda$ should be considerably larger 
than the gap.
Having fixed the cutoff, we will express the coupling constant 
$H$ through the corresponding value of the BCS gap. 

For given model parameters and $\delta\mu$, the thermodynamic potential,
as defined above, depends on the gap function $\Delta$.
Our main goal is to find the most stable solution, i.e., the minimum
of $\Omega$ with respect to $\Delta$.
At a given periodicity of the crystal, this corresponds to minimizing
$\Omega$ with respect to the Fourier components $\Delta_{q_k}$, 
i.e., to finding the most favored solution of the coupled set of gap 
equations, \eq{eq:gap2}.
In addition, we should vary the periodic structure itself, i.e.,
the basis vectors of the reciprocal lattice.
Obviously, this is a very involved problem, which is beyond the scope
of the present paper.  

Therefore, as a first step, we restrict ourselves to one-dimensional
crystalline structures, i.e., to gap functions which vary periodically
in one spatial direction $\hat q$, but stay constant in the two 
spatial directions perpendicular to $\hat q$.
Moreover, we only consider real gap functions.
In spite of these restrictions, we find an interesting class of solutions,
which will be discussed in Sec.~\ref{sec:genreal} and thereafter.
First, however, we briefly discuss the BCS and the Fulde-Ferrell solutions
within our model in order to provide a basis and to make contact to the 
literature.

\subsection{BCS phase}
\label{sec:BCS}

The BCS solution corresponds to the limit of spatially homogeneous 
condensates. In this case the Brillouin zone is the entire three-dimensional
space and we have
\beq
    \Delta_{q_k} \,=\, \Delta\,\delta_{q_k,0}~. 
\eeq
Therefore, the effective Hamiltonian \eq{HDm} is diagonal in momentum space,
and we obtain the eigenvalues
\beq
    |E_\pm(\vec k)| = |\sqrt{(k -\bar\mu)^2 + |\Delta|^2} 
    \pm \delta\mu|~. 
\eeq
Inserting this into \eqs{Omdmu} and (\ref{Om0dmuT0})
we recover the well-known result that
for $\delta\mu < |\Delta|$ the {\it unregularized} part of
the thermodynamic potential, 
\beq
\Omega_\mathit{unreg}(\delta\mu) = -4
\int\frac{d^3k}{(2 \pi)^3}
\sqrt{(k -\bar\mu)^2 + |\Delta|^2}
+ \frac{|\Delta|^2}{4H}~,
\eeq
is independent of $\delta\mu$ \cite{Bedaque:1999nu}.
When the Pauli-Villars regulators are included, this does no longer hold
exactly.
As a consequence, the value of $\Delta$ which solves the gap equation
$\partial\Omega/\partial\Delta^* = 0$ 
is weakly $\delta\mu$ dependent.
For this reason we {\it define} $\Delta_\mathit{BCS}$ to be the gap at $\delta\mu = 0$. 
For $\Delta_\mathit{BCS} = 80$~MeV and $\Lambda = 400$~MeV,
which will be our standard choice of parameters, 
we then find that $\Delta$ increases by about five percent, when $\delta\mu$ is 
varied between 0 and 0.8~$\Delta_\mathit{BCS}$. (Higher values of
$\delta\mu$ are irrelevant for the BCS phase, as we will see below.) 
For smaller gaps or larger values of the cutoff the effect is even smaller.




When $\delta\mu$ is increased, the BCS phase eventually becomes unfavored
against the normal phase. The corresponding phase transition is first
order. In the weak-coupling limit it occurs at 
$\delta\mu = \Delta_{BCS}/\sqrt{2}$, as shown already in 1962 by 
Chandrasekhar and Clogston \cite{ClCh1962}.
For stronger couplings, the BCS phase can sustain somewhat larger values
of $\delta\mu$.
We will come back to this in the following subsection.




\subsection{Fulde-Ferrell solutions}
\label{sec:FF}

The FF phase corresponds to the case that the gap function is a single
plane wave,
\beq
    \Delta(x) \,=\, \Delta\,e^{2i \vec q \cdot \vec x}~. 
\label{FFansatz}
\eeq
This means, the $B.Z.$ is infinite in the directions perpendicular to
$\vec q$, but finite in $\vec q$-direction with length $2|\vec q|$.
The momentum space representation of the gap matrix is given by
\beq
    \Delta_{q_k} \,=\, \Delta\,\delta_{\vec q_k,2\vec q}~.
\eeq
Inserting this into \eq{HDm} one finds that the effective Hamiltonian
is still block diagonal, however with shifted blocks
where two momenta are coupled.
These blocks can be written in the form
\beq
\left(
\begin{array}{cc}
k_+ -\bar\mu - \delta\mu &  \Delta
\\
\Delta^* & - k_- + \bar\mu - \delta\mu
\end{array}
\right)~,
\label{FFmatrix}
\eeq
with $k_\pm = |\vec k \pm \vec q|$.
Moreover, the sum over all blocks simply amounts to extending the 
$\vec k$-integration to the entire space. 
Hence, the thermodynamic potential reads
\beq
\Omega(\delta\mu) = -2
\int\frac{d^3k}{(2 \pi)^3}
\sum_{\lambda=\pm} \left(
|E_{\lambda}(\vec k)| \; + \; \mathit{reg.} \right)
    \,+\, \frac{|\Delta|^2}{4H}\,,
\label{OmegaFF}
\eeq
where
\beq
    E_{\pm}(\vec k) = \frac{k_+ - k_-}{2} 
    \pm\sqrt{ \left(\frac{k_+ + k_-}{2}-\bar\mu\right)^2 + |\Delta|^2} 
    -\delta\mu
\label{FFeigenvalues}
\eeq
are the eigenvalues of the matrix (\ref{FFmatrix}).
Apart from the Pauli-Villars regulators in \eq{OmegaFF}, this 
is again a standard result.

In order to find the most favored FF solution, we must minimize 
the thermodynamic potential with respect to $|\Delta|$ 
and $|\vec q|$.
Typically, one finds that $|\vec q|$ is of the order 
of $0.9\,\Delta_\mathit{BCS}$ while $|\Delta|$ is considerably smaller than 
$\Delta_\mathit{BCS}$.\footnote{In the weak-coupling limit,
$|\vec q| = 0.906\,\Delta_\mathit{BCS}$ and 
$|\Delta|= 0.23\,\Delta_\mathit{BCS}$ at the Chandrasekhar-Clogston
point \cite{Fulde:1964zz,LO64}.}
Numerical examples for $\Delta_\mathit{BCS} = 80$~MeV and 
$\Lambda = 400$~MeV will be discussed in the context of Figs.~\ref{fig:qdmu} and 
\ref{fig:Ddmu}.

\begin{figure}[ht]
\begin{center}
 \includegraphics[width=\linewidth]{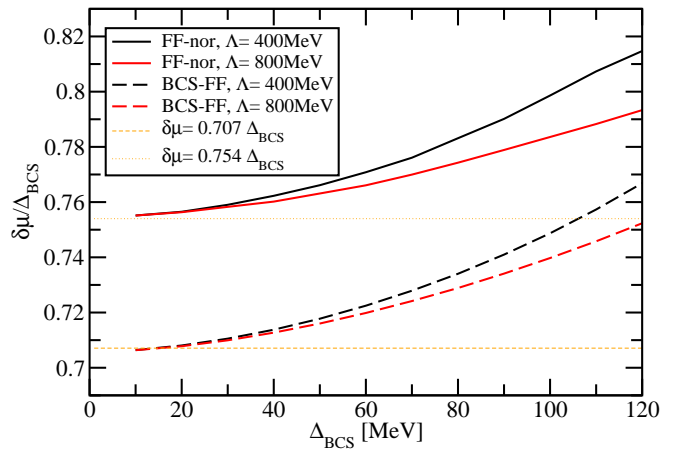}
 \caption{Stability window in $\delta\mu$ for the FF phase in 
 competition with the BCS phase and the normal phase as a function of
 the coupling, parameterized by $\Delta_\mathit{BCS}$.
 The BCS phase is favored below the dashed lines, the normal phase
 is favored above the solid lines.  
 Results for two different cutoffs are shown.
 The thin horizontal lines indicate the weak-coupling limits.
}
\label{fig:fish400}
\end{center}
\end{figure}

In competition with the normal phase and the BCS phase, this optimal
FF solution is favored only in a small window in $\delta\mu$. 
This is shown in Fig.~\ref{fig:fish400}, where the phase boundaries
are displayed as functions of $\Delta_\mathit{BCS}$ for
two different values of the cutoff. 
The phase transitions are of first order for the BCS-FF transition
(dashed lines) and of second order for the FF-normal transition (solid lines). 
Indeed, in the weak-coupling limit, one expects a first-order phase 
transition from the BCS phase to the FF phase near the Chandrasekhar-Clogston 
value $\delta\mu = \Delta_\mathit{BCS}/\sqrt{2}$ followed by the second-order
phase transition to the normal phase at $\delta\mu = 0.754\,\Delta_\mathit{BCS}$
\cite{Fulde:1964zz,LO64}. 
These values are indicated in the figure by the thin horizontal  lines.
Obviously, the model results come close to these limits for 
$\Delta_\mathit{BCS} \rightarrow 0$, whereas for stronger interactions
we again find deviations.

The BCS-FF phase boundaries are almost identical with
the corresponding BCS-normal boundaries 
because the free energy difference between BCS phase and 
normal phase is a much steeper function of $\delta\mu$ than the free 
energy difference between FF phase and normal phase
(see Fig.~\ref{fig:Omegadmu}).
This is a standard result, which was also found, e.g., in
Ref.~\cite{Alford:2000ze} where a different regularization scheme 
was used.

The FF-normal phase boundary, on the other hand, behaves rather differently
from the result of Ref.~\cite{Alford:2000ze}. There, it was found that 
the critical $\delta\mu$ decreases with increasing $\Delta_\mathit{BCS}$. 
As a consequence, the BCS-FF boundary and the FF-normal boundary intersect 
at some $\Delta_\mathit{BCS} \sim 100$~MeV, and there is no stable FF 
solution at stronger couplings. 
In contrast to this, we find that the FF-normal phase boundary
runs almost parallel to the BCS-FF phase boundary so that the 
width of the FF window stays approximately constant.
This means, 
as long as we do not consider other inhomogeneous phases, 
the regime of stable FF solutions extends to very large
values of $\Delta_\mathit{BCS}$ in our regularization scheme.


\subsection{General solutions for real one-dimensional gap functions}
\label{sec:genreal}

We now turn to the main part of this work, being the analysis of ground states
when allowing for a real one-dimensional gap function as the order parameter.
For simplicity, we will often call these solutions ``general solutions'',
although there should be, of course, even more general solutions with
higher-dimensional or complex gap functions. 

In the first step we limit ourself to periodic gap functions of the form
\bea
\Delta(x)
&=&
\sum_{k}
\Delta_{\vec{q},k}
e^{2ik\vec{q}\cdot\vec{x}}
\,.
\label{1Dansatz}
\eea
For each period given through $\vert\vec{q}\vert$ we will then minimize the
thermodynamic potential with respect to the Fourier components
$\Delta_{\vec{q},k}$ and afterwards with respect to $\vert\vec{q}\vert$.

Comparing \eq{1Dansatz} with \eqs{Deltaq} and (\ref{qk}), we find that
$\vec q_k = 2k \vec q$. The (admittedly somewhat unnatural) factor of 2
was introduced to give $\vec q$ the same meaning as in the FF ansatz,
\eq{FFansatz}, where this factor is the standard convention. 

Again the $B.Z.$ is infinite in the directions perpendicular to $\vec{q}$ and
finite in the $\vec{q}$-direction with length $2\vert\vec{q}\vert$.
Without loss of generality, we can assume that $\vec q$ points in 
z-direction, $\vec q = q\vec e_z$. 
Then one period corresponds to $z = \frac{\pi}{q}$.

The restriction to gap functions which are real in coordinate space
means that the Fourier components satisfy the relation 
$\Delta_{\vec{q},k} = \Delta^*_{\vec{q},-k}$. 
Moreover, without loss of generality,
we can always choose the origin to be located at a maximum
of the gap function. $\Delta(z)$ is then an even\footnote{In our 
numerical analysis, we only found solutions which are symmetric under
reflections at a plane perpendicular to $\vec q$ going through a 
maximum or minimum. So far, we cannot exclude that other solutions 
exist which do not have this symmetry. In this case $\Delta(z)$ would
of course not be an even function. \label{note}} 
function, and the Fourier components are real.


\begin{figure}[ht]
\begin{center}
 \includegraphics[width=\linewidth]{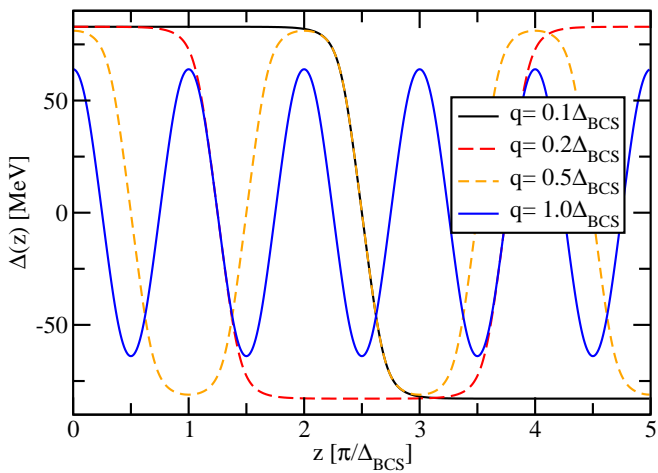}
 \caption{The gap function in coordinate space 
  at $\delta\mu=0.7\Delta_{BCS}$ for different fixed
  values of $q$.}
\label{fig:Dzdmu700}
\end{center}
\end{figure}


In the following we fix the model parameters to 
$\Lambda = 400$~MeV and a coupling strength corresponding to 
$\Delta_\mathit{BCS} = 80$~MeV. As before, we consider $T=0$ and
$\bar\mu = 400$~MeV.

In Fig.~\ref{fig:Dzdmu700} we present examples of one-dimensional
gap functions $\Delta(z)$ we obtained by minimizing the thermodynamic 
potential at $\delta\mu=0.7\Delta_\mathit{BCS}$ for different fixed periods.
For convenience, $z$ is measured in units of $\pi/\Delta_\mathit{BCS}$
so that one period is given by $(q/\Delta_\mathit{BCS})^{-1}$.
At $q \sim \Delta_{BCS}$ the gap function appears to be sinusoidal. 
For larger periods, however, a new feature becomes apparent: 
the formation of a soliton lattice.
Especially for $q=0.1\,\Delta_\mathit{BCS}$, we see that the gap function 
stays nearly constant at $\pm\Delta_{BCS}$ for about one half-period 
and then changes its sign in a relatively small interval.
The $q=0.2\,\Delta_\mathit{BCS}$ solution behaves qualitatively similar, 
but has a shorter plateau. Remarkably, the shape of the two functions 
is almost identical in the transition region where the gap functions 
change sign. This remains even true for the $q=0.5\,\Delta_\mathit{BCS}$ 
solution, which is kind of an extreme case with no plateau and only 
transition regions. We may thus interprete these transition regions
as very weakly interacting solitons, which are almost unaffected by the
presence of the neighboring (anti-) solitons as long as they do not
overlap. 

These features will be discussed in more detail in Sec.~\ref{sec:compto1d}.
The main result is that the gap functions are characterized by two 
independent scales. 
The first scale, $q$, determines the period of the lattice and thereby the
distance between the solitons. The second scale, $\Delta_\mathit{BCS}$,
is not only the amplitude but also determines the shape of the solitons,
which is practically independent of $q$. 
In fact, even for the sinusoidal solution at $q = \Delta_\mathit{BCS}$, 
the slope at the zero crossing is almost the same as for the 
$q = 0.1\,\Delta_\mathit{BCS}$ solution, although the period differs by
one order of magnitude.


\begin{figure}[ht]
\begin{center}
 \includegraphics[width=\linewidth]{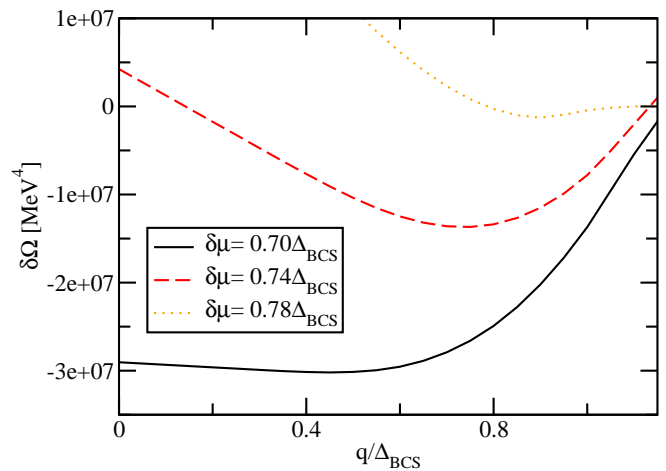}
 \caption{Difference between the thermodynamic potentials of the 
          general inhomogeneous phase and the normal phase as a 
          function of $q$ for different values of $\delta\mu$.}
\label{fig:Omegaq70}
\end{center}
\end{figure}


For each $\delta\mu$, 
with the solutions for the different chosen values of $q$ at hand,  
we now have to minimize the thermodynamic potential in $q$
in order to determine the energetically preferred ground state. 
This is illustrated in Fig.~\ref{fig:Omegaq70}, where 
the difference $\delta\Omega$ between the thermodynamic potential 
of the inhomogeneous phase and the normal phase is displayed
as a function of $q$ for three different values of $\delta\mu$.
We observe that the preferred value of $q$ rises from small
values towards $\Delta_{BCS}$ when increasing $\delta\mu$. 
Also the shape of the potential changes. 

We want to point out that the BCS phase is included in this plot as
$q=0$.
It is interesting to note that the inhomogeneous phase
is the preferred phase already at $\delta\mu = 0.70\,\Delta_\mathit{BCS}$,
i.e., significantly below the transition 
between the BCS phase and the FF phase
(cf.~Fig. \ref{fig:fish400}).


\begin{figure}[ht]
\begin{center}
 \includegraphics[width=\linewidth]{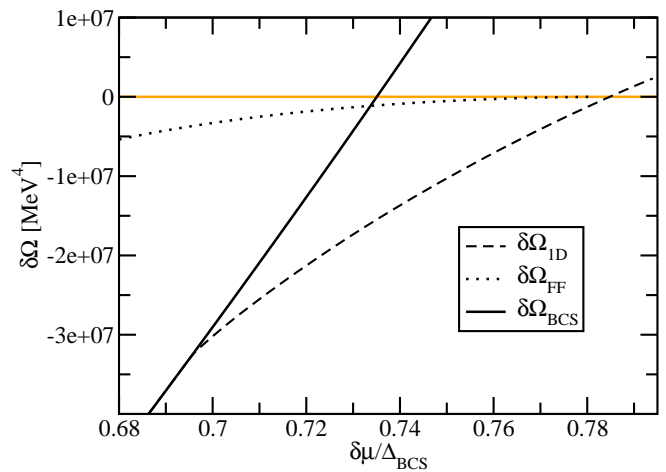}
 \caption{Difference between the thermodynamic potentials of different
          phases and the normal phase as functions of $\delta\mu$:
          BCS phase (solid line), general inhomogeneous phase 
          (dashed line), and FF phase (dotted line).}
\label{fig:Omegadmu}
\end{center}
\end{figure}


This is also seen in Fig.~\ref{fig:Omegadmu}, where $\delta\Omega$
is displayed as a function of $\delta\mu$ for the general solution
in comparison with the BCS phase and the FF phase. 
The shown results for the inhomogeneous phases correspond to the 
preferred values of $q$ at each $\delta\mu$.
We find that the LOFF window, i.e., the interval of $\delta\mu$ for which 
a general inhomogeneous ground state is energetically favored, has almost 
doubled compared to the FF window.
This is due to the fact that the interval is expanded towards smaller values
of $\delta\mu$, the upper end is almost unchanged.
Related to this, the free energy gain of the general solution is much 
larger than that of the FF phase. 

However, the most striking features which are visible in 
Fig.~\ref{fig:Omegadmu} are the orders of the phase transitions.
Whereas for the FF solutions, the phase transition to the BCS phase
is first order and to the normal phase is second order, 
it is just the other way around for the general inhomogeneous solutions: 
Here we find a first-order phase transition to the normal phase, 
whereas the transition to the BCS phase at the lower end appears to be 
continuous.


\begin{figure}[ht]
\begin{center}
 \includegraphics[width=\linewidth]{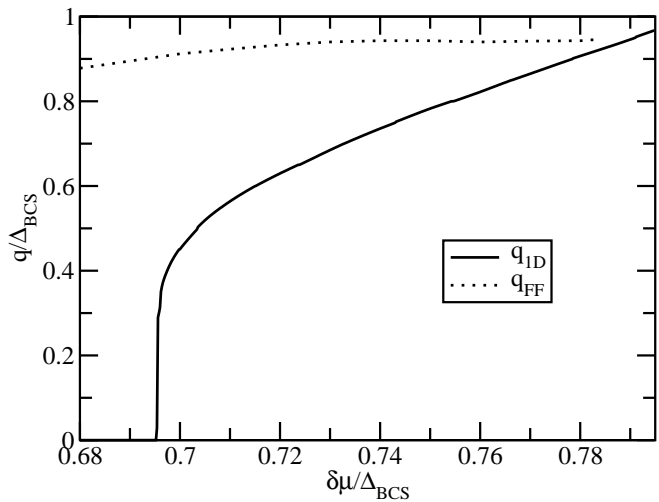}
 \caption{The energetically preferred value of $q$ in the general
 inhomogeneous superconducting phase (solid) 
 and for the FF solutions (dotted) as functions of 
 $\delta\mu$. The solution for the FF phase stops at the transition to the
 normal phase.
}
\label{fig:qdmu}
\end{center}
\end{figure}


The latter is possible because our space of possible solutions allows 
for a natural connection between these phases via the formation of 
a soliton lattice.
This becomes more clear in Fig.~\ref{fig:qdmu} where the energetically 
favored values of $q$ are displayed as functions of $\delta\mu$. 
The solid line and the dotted line correspond to the general inhomogeneous
phase and to the FF phase, respectively.
In the FF phase, $q$ is always of the order of $\Delta_\mathit{BCS}$.
Similar values of $q$ are also found for the general solutions in the
upper part of the LOFF window, $\delta\mu\gtrsim 0.75\Delta_\mathit{BCS}$.
With decreasing $\delta\mu$, however, the preferred $q$ decreases
to arbitrarily small values and eventually goes to zero at
$\delta\mu\approx 0.695\Delta_\mathit{BCS}$.

We thus arrive at the following picture:
With lowering $\delta\mu$, the period of the gap function increases
and we eventually obtain a soliton lattice
with increasing distance between the solitons, i.e., 
with constant plateaus of increasing 
length, see Fig.~\ref{fig:Dzdmu700}.
At the critical point, the length of the plateaus diverges,
and the inhomogeneous phase is continuously connected to the BCS phase.
Also notice that, although the transition is continuous, the slope of
the function $q(\delta\mu)$ changes dramatically when 
$q/\Delta_\mathit{BCS}$ comes to the order of 0.5. 
As we have seen in Fig.~\ref{fig:Dzdmu700}, this is just
the regime, where the more sinusoidal solutions go over into a soliton 
lattice.


\begin{figure}[ht]
\begin{center}
 \includegraphics[width=\linewidth]{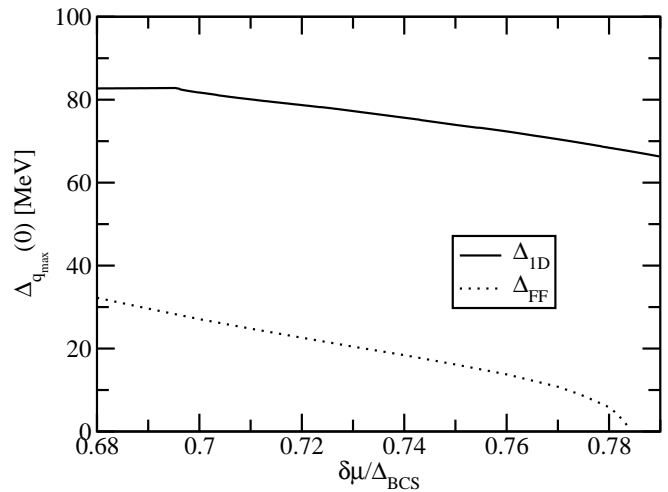}
 \caption{The amplitude of the gap function $\Delta(z)$ at the
 energetically preferred value of $q$ at given $\delta\mu$ for the general
 inhomogeneous phase (solid line) and the FF phase (dotted line).}
\label{fig:Ddmu}
\end{center}
\end{figure}


In Fig.~\ref{fig:Ddmu} the amplitude of the general inhomogeneous
gap function is displayed as a function of $\delta\mu$ (solid line).
Since the transition to the BCS phase is continuous, the amplitude
becomes equal to the BCS gap at the lower end of the window.\footnote{
We remind that, because of the regularization terms,
the BCS gap at $\delta\mu > 0$ is slightly larger than $\Delta_\mathit{BCS}$.} 
With increasing $\delta\mu$, the amplitude decreases, but it 
remains of the order of $\Delta_\mathit{BCS}$ in the entire interval.
For comparison, we also show  the amplitude of the FF solution (dotted line).
In contrast to the general solution it is always considerably smaller than 
$\Delta_\mathit{BCS}$ and goes to zero at the transition point to the 
normal phase. 
Thus, in agreement with our conclusions from Fig.~\ref{fig:Omegadmu}, 
the transition to the normal phase is of second order for the FF phase, 
but first order for the general inhomogeneous phase.
The fact that the amplitude of the general solution is considerably
bigger than the amplitude in the FF phase also reflects our earlier observation
that the general solution is energetically much more favored (see
Fig.~\ref{fig:Omegadmu}).


\begin{figure}[ht]
\begin{center}
 \includegraphics[width=\linewidth]{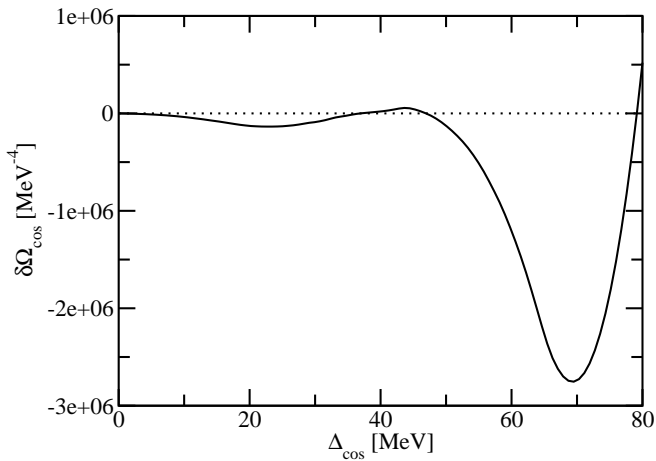}
 \caption{Difference between the thermodynamic potential of an inhomogeneous
   phase with a sinusoidal gap function and the normal phase as a function of
   the magnitude of the gap function. Here we have chosen $q=0.9\Delta_{BCS}$
   and $\delta\mu=0.775\Delta_{BCS}$.
}
\label{fig:OmegaGL}
\end{center}
\end{figure}


At first sight, a first-order phase transition from the inhomogeneous phase to
the normal phase seems to be in contradiction with Ginzburg-Landau
investigations. 
As already discussed in the context of
Figs.~\ref{fig:Dzdmu700} and \ref{fig:qdmu} and as will be detailed in
section~\ref{sec:compto1d}, the shape of the gap function becomes more and more
sinusoidal with increasing $\delta\mu$, i.e., increasing $q$.
A sinusoidal (or so-called antipodal) gap function has been investigated in a
Ginzburg-Landau analysis, showing that the transition from the inhomogeneous
phase to the normal phase is of second 
order~\cite{Bowers:2002xr,Casalbuoni:2003wh}.
Since in the vicinity of a second-order transition
the Ginzburg-Landau approximation is expected to converge to the exact
mean-field result, an explanation of this discrepancy is needed.

To clarify this issue, we determine the thermodynamic potential for a gap 
function
\bea
\Delta(z)
&=&
\Delta_{\mathrm{cos}}\cos(2 q z)
\eea
with fixed $q$, as being used in the Ginzburg-Landau analysis.
An example is shown in Fig.~\ref{fig:OmegaGL}, which corresponds to
$q=0.9\,\Delta_{BCS}$ and $\delta\mu=0.775\,\Delta_{BCS}$.
We see that the thermodynamic potential has two minima in this subset of
possible solutions.
The more shallow one at lower magnitudes of $\Delta_{\mathrm{cos}}$ may be
described by the Ginzburg-Landau analysis.
However, in order to get the second minimum, one needs at least to
include terms of the order $\Delta_{\mathrm{cos}}^8$, whereas in 
Refs.~\cite{Bowers:2002xr,Casalbuoni:2003wh} at most terms of the order
$\Delta_{\mathrm{cos}}^6$ were included.
Moreover, investigations in one spatial dimension have revealed
that a gradient expansion of the thermodynamic potential may even not
exist at all~\cite{Waxman1992}.
Therefore it is unclear whether the second minimum is determinable
in a Ginzburg-Landau approach, even if higher powers in the order
parameter are included.

\subsection{Quasiparticle spectrum}
\label{sec:spectrum}

Next we want to discuss the excitation spectrum of the quasiparticles and 
its impact on the shape of the energetically preferred gap functions.
For a given vector $\vec k$ in the $B.Z.$ we have a discrete eigenvalue 
spectrum of the associated Hamiltonian ${\cal H}_{\Delta,\delta\mu}(\vec k)$,
Eq.~(\ref{HDm}), which depends on $\vec k$ via its allowed momenta,
see Eqs.~(\ref{pmpn}-\ref{Hkmn}). 
As the inhomogeneous gap functions break the rotational symmetry of the
system, this eigenvalue spectrum does not only depend on the modulus
of $\vec k$, but also on its direction.
Therefore, we will present the spectrum by showing selected slices 
through the $B.Z$.

We note that the eigenvalue spectrum of ${\cal H}_{\Delta,\delta\mu}$
depends on $\delta\mu$ in two ways: 
First, there is an explicit $\delta\mu$ term in the
diagonal Nambu-Gor'kov components and, second, there is an implicit
$\delta\mu$ dependence through the $\delta\mu$ dependence of the
gap functions. 
For fixed gap functions, the explicit $\delta\mu$ terms simply shift 
the eigenvalues of ${\cal H}_{\Delta,0}$ by $-\delta\mu$.
We will therefore take this trivial effect out and show
the eigenvalue spectrum of ${\cal H}_{\Delta,0}$, which we will denote 
by $E_\lambda^{(0)}(\vec k)$. This spectrum 
then still depends on $\delta\mu$ through the gap functions.
The gap functions, in turn, depend on $\delta\mu$ mainly
through the variation of $|\vec q|$ (see Figs.~\ref{fig:Dzdmu700}
and \ref{fig:qdmu}). Therefore, we present the spectra at fixed 
values of $|\vec q|$. As we will see below, the remaining
$\delta\mu$ dependence is very weak, i.e., the spectra are quite
generic.

\begin{figure*}[ht]
  \begin{tabular}{cc}
    \includegraphics[width=0.5\linewidth]{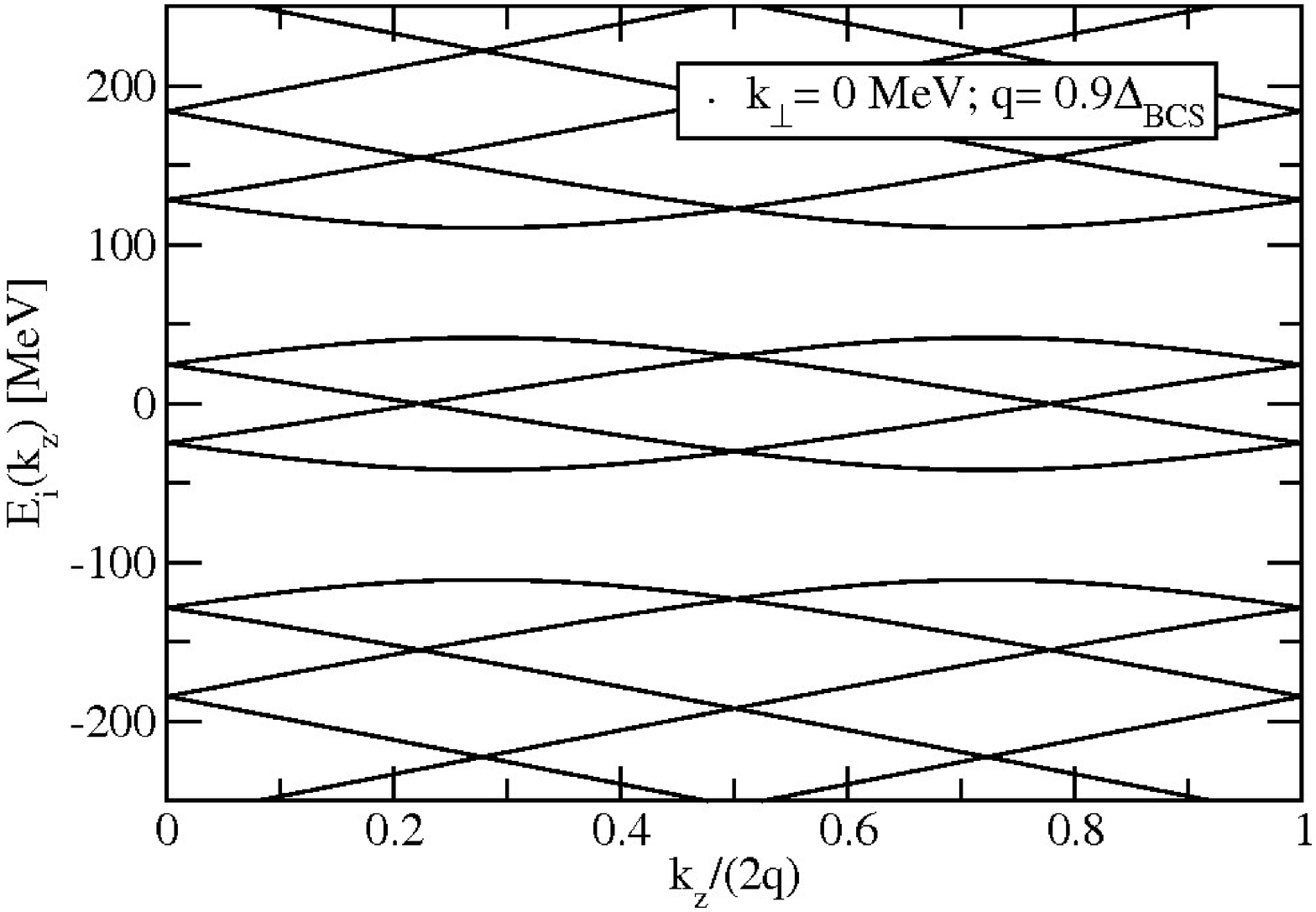}   &
    \includegraphics[width=0.5\linewidth]{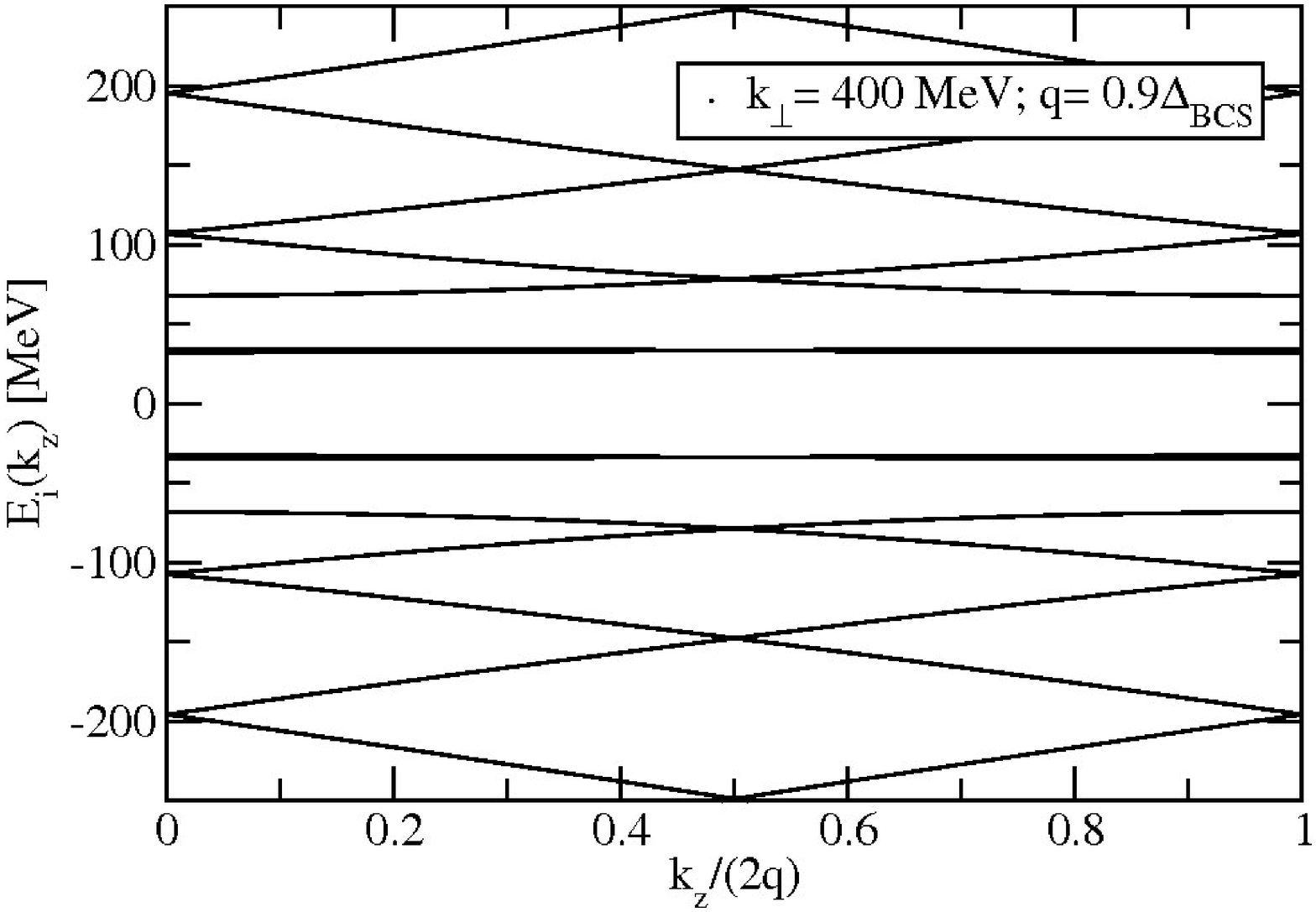}
  \end{tabular}
  \caption{Excitation spectrum in the $B.Z.$ as a function of the momentum
    $k_z$ parallel to $\vec{q}$ for fixed perpendicular momentum
    $k_\perp =0$ (left panel) and $k_\perp =400\,\mathrm{MeV}$ (right
    panel). Shown are the eigenvalues $E_\lambda^{(0)}(\vec k)$
    of ${\cal H}_{\Delta,\delta\mu=0}$
    for the gap functions of the energetically preferred ground state at fixed
    $\vert\vec{q}\vert=0.9\,\Delta_{BCS}$ and $\delta\mu=0.7\,\Delta_{BCS}$.
    The spectrum $E_\lambda(\vec k)$ of ${\cal H}_\mathit{HDE}$ is obtained by 
    shifting the lines by $\delta\mu$ and $-\delta\mu$.}
\label{fig:brio}
\end{figure*}

In Fig.~\ref{fig:brio} we show two examples of the excitation spectrum
at a fixed value of the momentum $k_\perp$ perpendicular to $\vec{q}$
as a function of the momentum $k_z$ along the direction of $\vec{q}$.
For a real gap function, the eigenvalue spectrum possesses two 
symmetries: First, as shown in Appendix~\ref{app:symevs}, the
eigenvalues $E_\lambda^{(0)}$ at given momentum $\vec{k}$ in the $B.Z.$ appear
in pairs $(E_\lambda^{(0)},-E_\lambda^{(0)})$.
Second, the eigenvalue spectrum at $k_z$ and $2\vert\vec{q}\vert-k_z$
coincide.
This is related to the fact that the gap functions are even functions 
in $z$, i.e., parity is unbroken by the condensate.\footnote{
This statement holds for our particular choice of the coordinate system.
More generally, the gap functions are symmetric under reflection at a 
plane perpendicular to $\vec q$ going through a maximum or minimum.
See also footnote~\ref{note}.}


In the left panel of Fig.~\ref{fig:brio} we show the spectrum 
for $k_\perp = 0$.
We see that there are four low-lying excitations with free energies well 
below $\Delta_\mathit{BCS}$. 
These modes are related to each other by the symmetries discussed above,
i.e., there is in fact only one non-trivial solution at low energies.
This feature is also known from analytical investigations in one spatial 
dimension and is related to the soliton~\cite{horovitz1981}.
In addition to these modes, there are (infinitely many) other excitations
with higher free energies. These higher-lying modes are clearly separated
from the low-lying ones, i.e., there is a gap in the excitation spectrum,
at least as long as we keep $k_\perp$ fixed.\footnote{We remind that here
we are discussing the eigenvalue spectrum $E_\lambda^{(0)}(\vec k)$
of ${\cal H}_{\Delta,\delta\mu=0}$.
The true spectrum of ${\cal H}_\mathit{HDE}$ is obtained by shifting the results
by $\delta\mu$ upwards {\it and} downwards, see \eq{HHDE}. The gap may then 
disappear in some cases.
This is, however, irrelevant for the discussion later on in this section,
which is based on the existence of a gap in the spectrum of 
${\cal H}_{\Delta,0}$.}
In contrast to the BCS phase, the gap does, however, not start at vanishing
free-energies.

The right panel of Fig.~\ref{fig:brio} corresponds to 
$k_\perp = \bar\mu = 400$~MeV. Most features of this spectrum are the same
as in the previous example. The main difference is the fact that now the
low-lying excitation does not change its sign when $k_z$ is varied.
As a consequence, the positive and the negative solutions do not cross
and there is an additional gap around vanishing free energies.

\begin{figure*}[ht]
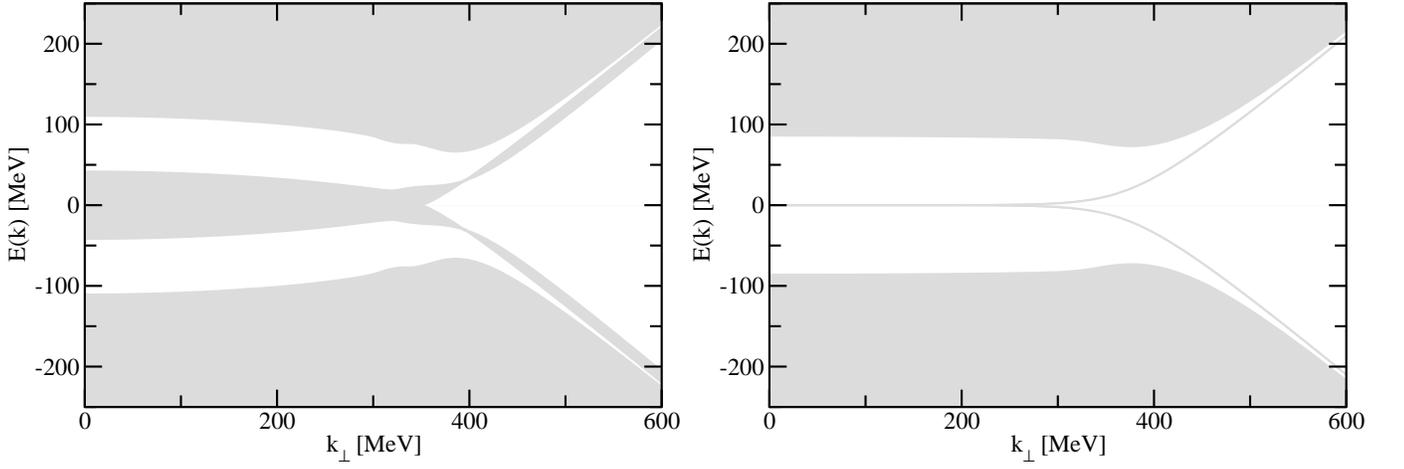

\begin{center}
  \begin{tabular}{cc}
    \includegraphics[width=0.5\linewidth]{brio2900v2.eps}&
    \includegraphics[width=0.5\linewidth]{brio2200v2.eps}
  \end{tabular}
  \caption{
    Superposition of the eigenvalue spectra $E_\lambda^{(0)}(\vec k)$
    of ${\cal H}_{\Delta,\delta\mu=0}$
    in the $B.Z.$ along the momentum
    $k_z$ parallel to $\vec{q}$ as functions of the perpendicular momentum
    $k_\perp$ (shaded areas). 
    The spectra have been obtained for  
    the gap functions of the energetically preferred ground state at 
    $\delta\mu=0.7\Delta_{BCS}$ for fixed periods 
    $\vert\vec{q}\vert=0.9\Delta_{BCS}$
    (left) and $\vert\vec{q}\vert=0.2\Delta_{BCS}$ (right).}
\label{fig:brio2}
\end{center}
\end{figure*}

The gap structure as a function of $k_\perp$ is visualized in 
Fig.~\ref{fig:brio2} for two examples. The shaded areas indicate the free 
energies which can be reached by at least one mode at the given value of 
$k_\perp$, when $k_z$ is varied over all possible values. 
Accordingly, the white areas correspond to the gapped regions.
We see again that such gaps exist in the excitation spectra when 
$k_\perp$ is kept fixed. There is, however, no gap when all 
values of $k_\perp$ are considered. 

The left panel of Fig.~\ref{fig:brio2} corresponds to 
$\vert\vec{q}\vert= 0.9\Delta_\mathit{BCS}$, i.e., to the same period as
the examples shown in Fig.~\ref{fig:brio}. Again, we see that
the low-lying modes are restricted to a single band 
around zero free energies for lower values of $k_\perp$,
wheres at higher values of $k_\perp$ this band splits into two.  
Comparing this with the right panel of Fig.~\ref{fig:brio2},
corresponding to $\vert\vec{q}\vert= 0.2\Delta_\mathit{BCS}$,
we see that these features remain qualitatively unchanged. 
However, we observe that the bands of the low-lying modes get
squeezed considerably when going from larger to smaller values of
$\vert\vec{q}\vert$, i.e., when separating the solitons by stretching 
the lattice. In fact, we expect that in the limit 
$\vert\vec{q}\vert\rightarrow 0$ and at low values of $k_\perp$,
the modes associated with the soliton are restricted to exactly
zero free energy, whereas the continuum starts at $|E| = \Delta_\mathit{BCS}$.


A gap in the complete excitation spectrum has important consequences for 
the dependence of the thermodynamic potential and the gap functions
on $\delta\mu$:
Consider the eigenvalues $E_\lambda^{(0)}(\vec k)$ of the 
unshifted Hamiltonian ${\cal H}_{\Delta,\delta\mu=0}(\vec k)$. 
Then, as pointed out above, the eigenvalues of ${\cal H}_{\Delta,\delta\mu}$
are simply given by $E_\lambda = E_\lambda^{(0)} - \delta\mu$.
Thus, since the $E_\lambda^{(0)}$ always come in pairs 
$(E_\lambda^{(0)},-E_\lambda^{(0)})$, we get from \eq{Om0dmuT0}
\beq
\Omega_{0}(\delta\mu) = 
-2\!
\int\limits_{B.Z.}\!\!\frac{d^3k}{(2 \pi)^3}\!\!
\sum_{E_\lambda^{(0)}>0} \! \left(
|E_\lambda^{(0)} - \delta\mu|  + |E_\lambda^{(0)} + \delta\mu|
\right)\,,
\eeq
where we have dropped the regularization terms for simplicity.
This can be written as 
\beq
\Omega_{0}(\delta\mu) = \Omega_{0}^{(1)} + \Omega_{0}^{(2)}(\delta\mu)~,
\eeq
with a $\delta\mu$ independent part
\beq
\Omega_{0}^{(1)} = -4\!
\int\limits_{B.Z.}\!\!\frac{d^3k}{(2 \pi)^3}\!
\sum_{E_\lambda^{(0)}>\delta\mu}
|E_\lambda^{(0)}(\vec k)|
\eeq
and a $\delta\mu$ dependent part
\begin{alignat}{1}
\Omega_{0}^{(2)}(\delta\mu)
&= -4\!
\int\limits_{B.Z.}\!\!\frac{d^3k}{(2 \pi)^3}\!\!\!
\sum_{0<E_\lambda^{(0)}<\delta\mu}\!\!\!\delta\mu
\nonumber \\
&\equiv
-4\,\delta\mu\!
\int\limits_{B.Z.}\!\!\frac{d^3k}{(2 \pi)^3} \, N_s(\vec k)
\;\equiv\; -4\,\delta\mu\,n_s~.
\end{alignat}
Here $N_s(\vec k)$ is the number of positive eigenvalues 
$E_\lambda^{(0)}(\vec k)$ smaller than $\delta\mu$. 

Now suppose, there was a complete gap in the spectrum in some
energy interval, so that $N_s(\vec k)$ is constant for all values
of $\vec k$, when $\delta\mu$ is varied within this gapped interval.
If $n_s = 0$, i.e., if $\delta\mu$ is smaller 
than the smallest positive eigenvalue, the thermodynamic potential and,
hence, the gap function are obviously $\delta\mu$ independent.
For $n_s > 0$, on the other hand, the thermodynamic potential is
$\delta\mu$ dependent.
However, even in this case, the {\it gap functions} are still 
$\delta\mu$ independent, when $\delta\mu$ is varied within the gapped 
interval. This can be inferred from the gap equations, 
$\frac{\delta\Omega}{\delta\Delta_{\vec q,k}} = 0$.
A $\delta\mu$ dependence of the gap function must then be due to the 
variation $\frac{\delta\Omega_0^{(2)}}{\delta\Delta_{\vec q,k}}$,
because all other terms are $\delta\mu$ independent. 
On the other hand, an infinitesimal variation of the gap function 
can only lead to infinitesimal changes in the eigenvalue spectra.
Therefore, if $\delta\mu$ lies in a gapped region, 
this variation will not change the numbers $N_s(\vec k)$,
which are, by definition, integer numbers. 
Consequently, $\frac{\delta\Omega_0^{(2)}}{\delta\Delta_{\vec q,k}} = 0$,
and the gap function is $\delta\mu$ independent.\footnote{In our model, 
this property is slightly spoiled by the regularization, as already
discussed for the BCS gap.} 


\begin{figure}[ht]
\begin{center}
 \includegraphics[width=\linewidth]{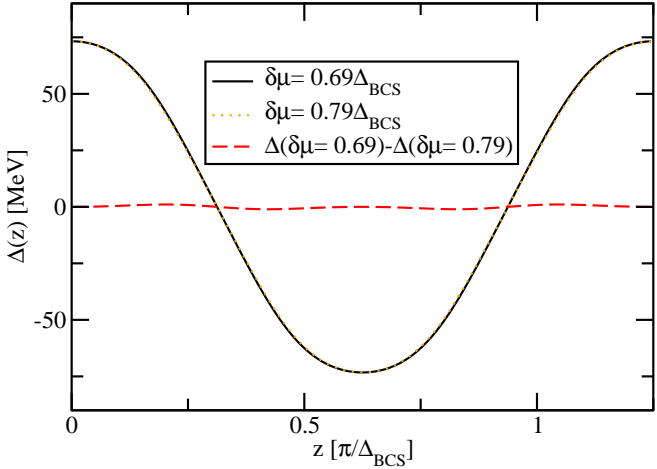}
 \caption{Gap functions at $\delta\mu=0.69\Delta_{BCS}$ and
   $\delta\mu=0.79\Delta_{BCS}$, and their difference
   for $\vert\vec{q}\vert=0.8\Delta_{BCS}$.
}
\label{fig:Deltadmu80}
\end{center}
\end{figure}


In our case the excitation spectrum is not completely gapped. Nevertheless,
when varying $\delta\mu$ in the vicinity of $0.7\,\Delta_{BCS}$, 
the only eigenvalues interfering
are those associated to the solitons and with momenta
$\vert\vec{k}\vert>\bar{\mu}$.
There number is however relatively small and they are outside  the Fermi ball.
We can therefore expect their influence on the value of the
thermodynamic potential and, hence, on the shape of the gap function to be
very small.

To illustrate this property we present the gap function at
$\delta\mu=0.69\,\Delta_{BCS}$ and
$\delta\mu=0.79\,\Delta_{BCS}$ for $\vert\vec{q}\vert=0.8\,\Delta_{BCS}$ in
Fig.~\ref{fig:Deltadmu80}.
The gap functions, though taken at the lower and upper end of the LOFF window,
appear almost identical.
We conclude that the shape of the gap functions in the LOFF window for given
$\vert\vec{q}\vert$ is almost independent of $\delta\mu$.

We also note that the spatially averaged density difference between
up and down quarks is given by
\beq
    \ave{\delta n} = \ave{n_u} - \ave{n_d} 
    = -\frac{\partial\Omega}{\partial\,\delta\mu}\,.
\eeq
We thus get from the above equations
\beq
\ave{\delta n} = 4 \left( n_s + 
    \delta\mu \, \frac{\partial n_s}{\partial\,\delta\mu} \right)\,,
\eeq 
where we have used that the implicit $\delta\mu$ dependence through
the gap functions drop out because of the gap equations.
Moreover, as we have seen before, $n_s$ is almost independent of $\delta\mu$, 
if $\delta\mu$ is varied within the nearly gapped region. Therefore the
second term on the right hand side is small and we find that 
the density difference is approximately proportional to the density
of solitonic states, 
\beq
    \ave{\delta n} \approx  4 n_s\,.  
\eeq
In particular, this implies that, unlike in the BCS phase, the density
difference does not vanish if low-lying solitons are present.
On the other hand, $\ave{\delta n}$ is almost $\delta\mu$ independent
when $\delta\mu$ is varied in the gapped regime.  

This gives rise to the very intuitive picture that the excess quarks are
sitting in the solitons, whereas in the BCS-like plateaus of the gap 
functions the densities of up and down quarks are nearly equal.
Here we should keep in mind that the period of the gap function and, hence,
the soliton density, was kept constant in the above discussion. 
This explains why we found $\ave{\delta n}$ to be $\delta\mu$ independent.
If we do not fix the period, but minimize the thermodynamic potential with
respect to $|\vec q|$, as in Fig.~\ref{fig:qdmu}, $\Omega$ is no longer
linear in $\delta\mu$, and $\ave{\delta n}$ becomes $\delta\mu$
dependent. In fact, we can turn this argument around
and explain the $\delta\mu$ dependence of $|\vec q|$ by the preference of 
the system to accommodate more excess quarks with increasing $\delta\mu$.

\subsection{Comparison with analytical results in one spatial dimension}
\label{sec:compto1d}

It is quite instructive to compare our results for the gap functions
in three spatial dimensions with the analytically known solutions of the 
mean-field problem in one spatial dimension.

In $1+1$-dimensions the mean-field problem can be approached in 
different ways and its first solution goes back to Peierls~\cite{peierls1955}.
In terms of inverse scattering theory, the selfconsistently determined gap
function has to generate a reflectionless (for a single bound state) or
finite-gap potential in the Hamiltonian~\cite{dashen1975,shei1976} leading to a single band in the excitation spectrum.
This is very similar to the feature displayed in Fig.~\ref{fig:brio} and
Fig.~\ref{fig:brio2} for small perpendicular momenta $k_\perp$.
Related results have also been obtained in the Gross-Neveu model~\cite{Thies} and,
recently, selfconsistent solutions have also been found for complex gap
functions~\cite{Basar:2008im}.

The case of a superconductor with two non-degenerate fermion species in
$1+1$-dimensions has been discussed in Ref.~\cite{machida1984}.
It was found that there is a transition at 
$\delta\mu=\frac{2}{\pi}\Delta_{BCS}$ from the BCS phase to the solitonic 
phase, which persists to any larger value of $\delta\mu$ at zero temperature.
A selfconsistent solution of the gap function in this case is given by
\bea
\label{eq:sn1d}
\Delta_{1+1}(x)
&=&
\kappa\sqrt{\nu}\;
\mathrm{sn}\big(\kappa(x-x_0);\nu\big)
\,,
\eea
where $\mathrm{sn}(\xi;\nu)$ is a Jacobi elliptic function with elliptic 
modulus $\nu^2$. 
The Jacobi elliptic function has the properties 
$\mathrm{sn}(0;\nu)= 0$, 
$\frac{\partial}{\partial\xi}\mathrm{sn}(\xi;\nu)|_{\xi=0}=1$,
and for $0\leq\nu<1$ it is periodic in $\xi$ with period $4K(\nu)$,
where $K(\nu)$ is the complete elliptic integral of the first kind.
For the limiting cases, we have $\mathrm{sn}(\xi;0) = \sin\xi$ and
$\mathrm{sn}(\xi;1) = \tanh\xi$, i.e.,
for $\nu\rightarrow 0$ the gap function becomes sinusoidal, whereas
in the limit $\nu\rightarrow 1$ we recover a single soliton. 

With \eq{eq:sn1d}, the ground state is obtained by minimizing the 
thermodynamic potential in $\kappa$ and $\nu$.
It is remarkable that -- apart from $x_0$, which is just an
arbitrary shift --
the gap functions can be characterized by only two parameters,
$\kappa$ and $\nu$,
which can be related to the period $4K(\nu)/\kappa$ and to the
slope $\kappa^2\sqrt{\nu}$ at the zero crossings.
In particular, the amplitude $\kappa\sqrt{\nu}$ is completely determined
by these two quantities.
Also note that $\kappa$ is just the slope at the zero crossings 
divided by the amplitude and we can therefore identify $2/\kappa$
with the soliton size. Moreover, for constant $\kappa$,
if $\nu$ is varied within an interval close to 1,
the period changes strongly, whereas the amplitude and the slope
stay nearly constant. This is very similar to our observation in 
Fig.~\ref{fig:Dzdmu700} that the amplitude and the shape of the 
transition region is almost independent of the period.

One may thus wonder, whether the one-dimensional gap functions in the 
$3+1$ dimensional system are given by similar functions.
Since, at least in the weakly coupled regime, all
dynamics is constrained to a close shell around the Fermi surface,
we could study the system on small patches at the Fermi surface, 
where the $3+1$-dimensional problem can be reduced to  
$1+1$-dimensional ones by applying a quasi-classical 
approximation~\cite{Andreev1964}.
The essential difference to the real $1+1$ dimensional case is the fact
that in $3+1$ dimensions the momenta of the paired quarks are in 
general not parallel to the wave vector $\vec q$ of the condensate.
Therefore the quasi $1+1$-dimensional problem needs to be solved for 
all directions~\cite{Waxman1994,Kosztin1998}.
However, projecting the wave-vector of the condensate on a given direction 
renders the relation between shape and period of the order parameter 
and its amplitude to depend on the direction.
For this reason, the $1+1$-dimensional approach for obtaining a 
selfconsistent solution cannot be extended trivially to $3+1$-dimensions.
 
On the other hand, we already observed in the discussion of 
Fig.~\ref{fig:Dzdmu700} that, just as in $1+1$ dimensions,
the gap functions seem to depend on two independent scales, 
$|\vec q|$ and $\Delta_{BCS}$, and that the latter determines both,
the amplitude and the soliton size.
Taking into account the complications discussed in the previous 
paragraph, this could be understood if the pairing is dominated 
by regions of the Fermi surface with a fixed azimuthal angle with
respect to $\vec q$.
This idea is also inspired by Ginzburg-Landau investigations where
the pairing mechanism of the fermions shows up more
explicitly~\cite{Bowers:2002xr}.
In particular for the FF phase at weak coupling, the pairing is 
concentrated around rings on the Fermi surface 
where the relative angle between $\vec k$ and $\vec q$ is given 
by $\cos\theta_\mathit{FF} \approx \frac{\delta\mu}{|\vec q|}$.
With this picture in mind we could imagine a similar regime of the Fermi ball
to dominate the pairing in the general inhomogeneous phase with the order
parameter varying in only one dimension.
We would then expect that the amplitude of the gap function is again
proportional to the inverse size of the soliton, but with a different 
proportionality constant than in the $1+1$-dimensional case.
Based on these arguments we suggest the fit
\bea
\label{eq:fitform}
\Delta_{\text{fit}}(z)
&=&
A\,\mathrm{sn}\big(\kappa(z-z_0);\nu\big)
\eea
as a parameterization of our numerical results.
To describe a gap function with period $\frac{\pi}{\vert\vec{q}\vert}$
and to comply with our convention of taking $\Delta(z)$ even with a maximum 
at $z=0$, we choose
\beq
A = \Delta(0)\,,
\quad
\kappa = \frac{4K(\nu)\vert\vec{q}\vert}{\pi}\,,
\quad
z_0 = -\frac{\pi}{4\vert\vec{q}\vert}\,.
\label{eq:fitpara}
\eeq
Then the only fit parameter left is $\nu$.
As a measure for the quality of the fit we consider the relative deviation
\bea
\frac{
  \Vert \Delta-\Delta_{\text{fit}}\Vert_2
}{
  \Vert\Delta_{\text{fit}}\Vert_2
}
&=&
\frac{\Vert \Delta-\Delta_{\text{fit}}\Vert_2}{\Delta(0)}
\left(\frac{\nu\,K(\nu)}{K(\nu)-E(\nu)}\right)^{\frac{1}{2}}
\,,
\eea
where $E(\nu)$ is the complete elliptic integral of the second kind
and $\Vert .\Vert_2$ is the $L^2$-norm.


\begin{figure}[ht]
\begin{center}
 \includegraphics[width=\linewidth]{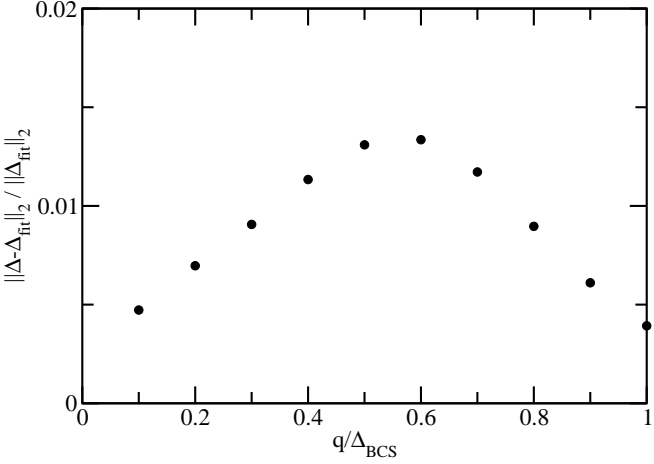}
 \caption{Relative error between the numerically determined gap function
          and its best fit by \eq{eq:fitform}
          as function of the period. Here we have chosen 
          $\delta\mu=0.7\Delta_{BCS}$.
}
\label{fig:Deltaerr70b}
\end{center}
\end{figure}



\begin{figure}[h]
\begin{center}
 \includegraphics[width=\linewidth]{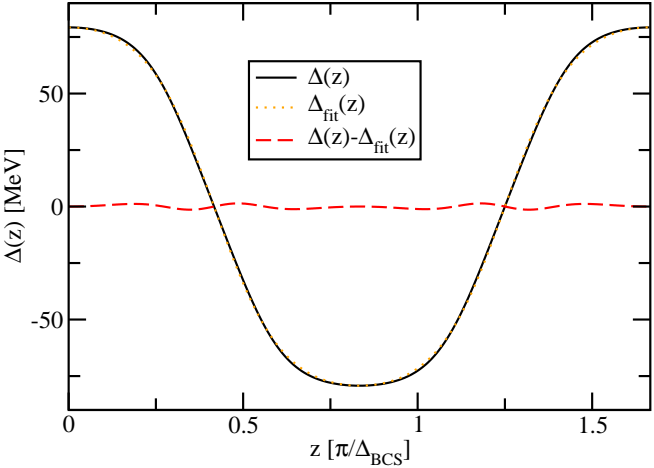}
 \caption{
   Numerical result for gap function $\Delta(z)$ compared to the best fit
   $\Delta_{\text{fit}}(z)$ at $\delta\mu=0.7\Delta_{BCS}$ and
   $\vert\vec{q}\vert=0.6\Delta_{BCS}$.
}
\label{fig:Deltafit60}
\end{center}
\end{figure}

In Fig.~\ref{fig:Deltaerr70b} we present the relative error of our numerically
obtained solutions compared to the best choice for $\nu$ in the
parameterization.
We have chosen $\delta\mu=0.7\Delta_{BCS}$. However, we remind that the gap
function is very insensitive under variation of $\delta\mu$, as pointed out in
the context of Fig.~\ref{fig:Deltadmu80}.
It turns out that gap function is described remarkably well by the
parameterization in Eq.~(\ref{eq:fitform}) with only one fit parameter.
The relative deviation in the $L^2$-norm is of the order of one
percent\footnote{In view of the involved numerics and the marginal deviation 
we even do not want to exclude that the functions are identical.}.
To illustrate this further, we present the ``worst'' case at
$\vert\vec{q}\vert=0.6\Delta_{BCS}$ in Fig.~\ref{fig:Deltafit60}.
The difference between the two functions is barely visible from plot.

\begin{figure}[h]
\begin{center}
 \includegraphics[width=\linewidth]{costhetav2.eps}
 \caption{$\cos\theta$ as defined in Eq.~(\ref{eq:costheta}) as function of
   $\vert\vec{q}\vert$ for $\delta\mu=0.7\Delta_{BCS}$.
}
\label{fig:costheta}
\end{center}
\end{figure}


\begin{figure}[h]
\begin{center}
 \includegraphics[width=\linewidth]{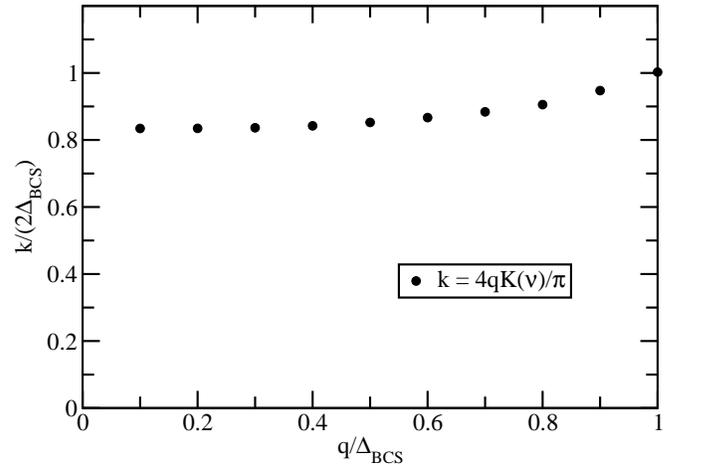}
 \caption{Inverse soliton size $\kappa$ as function of $\vert\vec{q}\vert$ for
   optimum fit to numerical result. $\delta\mu=0.7\Delta_{BCS}$.
}
\label{fig:kfitb}
\end{center}
\end{figure}

Having found the astonishing agreement between the analytical form inspired
from $1+1$-dimensional investigations and our numerical results, we are
naturally lead to the assumption that the underlying dominant pairing
mechanism is indeed dominated by fermions on the Fermi surface whose momenta 
have a fixed azimuthal angle measured from the direction of $\vec{q}$.
If this is true, we would expect that the amplitude of the gap function
is again proportional to $\kappa$, but with a different proportionality
constant as in \eq{eq:sn1d}. 

To establish this connection, which has so far been neglected in our 
parameterization, we assume that the pairing in directions $\hat{n}$ with a 
particular azimuthal angle $\theta$ to the direction of $\vec{q}$ is 
dominant. Furthermore, we assume that the gap function in $3+1$ dimensions 
is just given by the $1+1$ dimensional  result, \eq{eq:sn1d}, if we go 
along a line in such a direction, i.e.,
\beq
    \Delta_{3+1}(z) \= \Delta_{1+1}(x_\theta(z))\,,
\eeq
where $x_\theta(z) = \frac{z}{\cos\theta}$ is the coordinate in 
$\hat n$ direction for a given $z$
and
\beq
  \Delta_{1+1}(x_\theta) =
  \kappa_\theta\sqrt{\nu}\;
  \mathrm{sn}\big(\kappa_\theta(x_\theta-x_{\theta,0});\nu\big)
\eeq
corresponds to \eq{eq:sn1d}.
Hence, if we take $\kappa_\theta = \kappa\cos\theta$, we obtain
\beq
    \Delta_{3+1}(z) \= 
    \kappa\cos\theta\sqrt{\nu}\;
    \mathrm{sn}\big(\kappa(z-z_0);\nu\big)\,.
\eeq
Comparing this with our ansatz \eq{eq:fitform} and \eq{eq:fitpara}
we get
\beq
    \cos\theta = \frac{A}{\kappa\sqrt{\nu}} =
    \frac{\pi\Delta(0)}{4\sqrt{\nu}K(\nu)\vert\vec{q}\,\vert}
\,.
\label{eq:costheta}
\eeq
This quantity is shown in Fig.~\ref{fig:costheta}. It turns out that
our suggested picture breaks down at
$\vert\vec{q}\vert\gtrsim 0.9\Delta_{BCS}$, where $\cos\theta$ 
becomes larger than unity. 
This is related to the fact that the maximum of the gap function in our
numerical results remains large also when its shape becomes more and more
sinusoidal towards larger values for $\vert\vec{q}\vert$. In the limiting 
case we have $\nu\rightarrow 0$ and Eq.~(\ref{eq:costheta}) blows up.

On the other hand, we find that $\cos\theta$ is almost constant for
$\vert\vec{q}\vert\lesssim 0.6\,\Delta_{BCS}$ at a value of 
$\cos\theta\approx 0.62$. This value corresponds to an opening angle 
$2\theta \approx 100^\circ$, somewhat larger than the typical opening angle 
of $67^\circ$ in the FF phase. 
Note, however, that the situation in the FF phase is 
rather different. There the pairing can be understood most intuitively
by shifting two Fermi spheres with radii $\bar\mu + \delta\mu$ and
$\bar\mu - \delta\mu$ by $\vec q$ in opposite directions \cite{Alford:2000ze}. 
In weak coupling, the physically relevant regime is then constrained around 
the intersection of the two Fermi surfaces,
and one finds that $\cos\theta_\mathit{FF} \approx \frac{\delta\mu}{|\vec q|}$,
as mentioned before. 
This means that $\theta_\mathit{FF}$ is not constant as a function of 
$|\vec q|$ and only possible if $|\vec q| \gtrsim \delta\mu$.
However, since in the FF phase $|\vec q|$ by itself is almost 
constant $\sim 0.9\Delta_{BCS}$, one typically finds
$\cos\theta_\mathit{FF}\approx 0.83$.

For the general one-dimensional inhomogeneous phase, this is quite
different. In this case, as we have seen, $|\vec q|$ varies strongly and 
can become arbitrarily small.
Nevertheless, according to Fig.~\ref{fig:costheta}, the associated angle
is approximately constant, at least for $|\vec{q}|\lesssim 0.6\,\Delta_{BCS}$.
This regime, where FF pairing would not be possible at all,
agrees roughly with the realm of the soliton lattice, 
see Fig.~\ref{fig:Dzdmu700}. 
Here the gap functions are no longer dominated by the lowest
Fourier components $\Delta_{\vec q,\pm 1}$, see \eq{1Dansatz}, 
but higher harmonics are important as well.
In fact, as we have seen earlier, the scale $|\vec q|$ which
is related to the inverse distance of the solitons becomes
rather irrelevant for the dynamics, which is more closely related
to the inverse soliton size $\kappa$.  

The latter is displayed in Fig.~\ref{fig:kfitb}.
It is interesting to see that its value is comparable with the 
corresponding scale in the FF and indeed almost independent of the 
periodicity of the solitons. 
From the numbers which can be read off from the figure we find that
the soliton size $\frac{2}{\kappa}$ is about $0.3$ - $0.4$ in units
of $\frac{\pi}{\Delta_{BCS}}$. This is in good agreement with 
Fig.~\ref{fig:Dzdmu700}.

\section{Summary and conclusions}
\label{sec:summary}

We have studied inhomogeneous pairing in relativistic imbalanced Fermi 
systems within a model of the Nambu--Jona-Lasinio type.
The analysis was performed within mean-field approximation, but without
restriction to the Ginzburg-Landau approach.
In the set-up of the model we mainly focused on color superconducting 
phases, but the method is rather general and could be applied, e.g., to
condensed matter physics or cold atomic gases as well. 

In Sec.~\ref{sec:formalism} we developed the general formalism for
calculating the thermodynamic potential. 
We considered a certain class of gap functions, which are time 
independent and periodic in space.
The color and flavor structure of the gap functions was chosen 
according to the most important pairing patterns in dense QCD,
CFL and 2SC, but the extention to other phases is straight forward.

We pointed out that the regularization of the theory needs to be 
addressed with special care in order to get a consistent description of
homogeneous and inhomogeneous phases. 
For instance, a straight-forward generalization of the three-momentum 
cutoff scheme to inhomogeneous systems leads to undesired artifacts.
To avoid these problems we suggest a Pauli-Villars-like regularization
scheme, which can be derived via proper-time regularization.
In this scheme the divergencies are regulated by restricting the free 
energies of the quasiparticles, rather than their momenta. 
As a consequence, the results are rather insensitive to the choice of 
the cutoff parameter if the gap in the homogeneous phase is fixed. 
In particular, model independent results in the weak-coupling limit
are reproduced correctly. 

In Sec.~\ref{sec:results} we discuss first numerical results obtained
within this framework. 
To this end, we considered a simplified model with two fermion species 
at a chemical potential difference $\delta\mu$.
Basically, this corresponds to a restriction to the 2SC phase in
high-density approximation.
Moreover, in order to keep the numerical effort at a tractable level,
we restricted ourselves to real gap functions with general periodic 
structures in one dimension.
With this ansatz we found that the inhomogeneous solutions are favored
against homogeneous superconducting (BCS) and normal conducting phases 
in a $\delta\mu$-window which is about twice as wide as for a simple 
plane-wave ansatz (FF phase). The main effect is seen at the lower
end of this window, i.e., towards the boundary to the BCS phase.
In this region we observe the formation of a soliton lattice. 
With lowering $\delta\mu$ the distance between the solitons increases
and eventually diverges at the critical point. In this way the 
inhomogeneous phase is continuously connected to the BCS phase. 
On the other hand, at the upper end of the window where the inhomogeneous 
phase is favored, the gap functions are sinusoidal, and the transition 
to the normal conducting phase is of first order.

It is worth noting that similar investigations of inhomogeneous phases of
cold atoms in the unitary regime using a local density functional
approach~\cite{Bulgac:2008a,Bulgac:2008b} give exactly the opposite ordering
of the phase transitions, i.e., first order for the transition from the
homogeneous to the inhomogeneous phase and second order from the inhomogeneous
to the normal phase.
In our approach the behavior is however not unexpected: On the one hand side,
mean-field investigations in $1+1$-dimensions find the same
behavior~\cite{machida1984} for the transition from the homogeneous to the
inhomogeneous phase.
On the other hand, the phase transition near the
tricritical point at finite temperature can be analyzed in a generalized
Ginzburg-Landau approach~\cite{Buzdin:1997a,Houzet:1999a}.
The obtained results are perfectly consistent with ours.
At this point we would also like to mention that within the generalized
Ginzburg-Landau approach the energetically preferred inhomogeneous phase near
the tricritical point has an order parameter only varying in one
dimension, as in our investigation.
It is however not clear whether this persists to zero temperature.

We also studied the quasiparticle excitations in the inhomogeneous 
ground state as functions of the momentum $\vec k$ in the $B.Z.$ 
As the inhomogeneous gap functions break the rotational symmetry,
the spectra depend on both, the modulus and the direction of $\vec k$,
leading to an interesting band structure. 
We found that the spectrum is ``almost gapped'' with a few low-lying modes
which correspond to the solitons. As a consequence, the gap functions
are almost insensitive to variations of $\delta\mu$ if the period is
kept constant.

Finally, we compared our solutions for the one-dimensional gap functions
with the known analytical solutions of the corresponding 1+1 dimensional 
theory, i.e., Jacobi elliptic functions.
We found that we can achieve an excellent fit in 3+1 dimensions,
if we allow for one additional parameter. 
This can be interpreted as an effective 1+1 dimensional behavior 
which comes about if the pairing is dominated by fermions
on the Fermi surface with momenta at a fixed azimuthal angle
relative to $\vec q$.

The present paper is only a first step towards a more complete
description of inhomogeneous pairing in relativistic systems. 
In the future, these studies should be extended in several directions:

We have calculated the gap functions as functions of a given
chemical potential difference $\delta\mu$. It would be interesting to
see how this translates into density profiles of the different fermion
species. As argued at the end of section~\ref{sec:spectrum},
we expect that the density difference will be close to zero
at the constant plateaus of the gap functions and that it will be peaked
at the zero crossings, i.e., in the solitons.  
Having worked this out, one could apply this to describe more physical
situations, like globally neutral two-flavor quark matter in 
beta equilibrium or imbalanced atomic systems with fixed concentrations.

The analysis could be extended to finite temperature. In the 
context of color superconductivity one should also study other pairing 
patterns, like the CFL phase. Eventually,
all this should be combined to obtain a phase diagram where homogeneous
and inhomogeneous phases are treated on an equal footing. It would be
interesting to see whether this can cure the problem of chromomagnetic 
instabilities in the phase diagram of neutral quark matter. 

Of course, it would be desirable to relax the restriction to 
one-dimensional gap functions and to study two and three-dimensional
crystalline structures. 
In fact, earlier investigations of crystalline color superconductivity
revealed that three-dimensional crystals are the most favored 
inhomogeneous solutions 
\cite{Bowers:2002xr,Rajagopal:2006ig,Casalbuoni:2005zp}.
These analyses, however, were performed in Ginzburg-Landau
approximation, which turned out to be particularly problematic in the
two-flavor case \cite{Bowers:2002xr}.
Moreover, the crystal structures were restricted to superpositions of
a finite number of plane waves whose wave vectors all have the same
length. 
This means that, e.g., the non-sinusoidal one-dimensional solutions 
which we found close to the BCS regime were not included.
It is therefore not clear a priori which solutions are favored in a
more complete analysis with arbitrary one-, two-, and three-dimensional
periodic structures,
in particular also because a one-dimensional solution is expected near the
tricritical point~\cite{Buzdin:1997a,Houzet:1999a}.
In principle, this could be studied within our framework.
In practice, of course, the numerical solutions will become very involved,
and one has to see how far one can get.

\section*{Acknowledgments}
We thank G. Basar, G. Dunne, M. Forbes, H. Gies, M. Mannarelli and especially
K. Rajagopal for discussions and comments.
This work was supported in part by funds provided by the German Research
Foundation (DFG) under grant number Ni 1191/1-1, by the Helmholtz-University
Young Investigator Grant No VH-NG-332 and by the U.S. Department of
Energy (D.O.E.) under cooperative research agreement DE-FG0205ER41360.
\appendix

\section{Ginzburg-Landau expansion and requirements on regularization 
procedure}
\label{app:GL}

Starting from the mean-field thermodynamic potential, \eq{eq:menfieldOm},
we can derive a Ginzburg-Landau functional by expanding the potential
in powers of the pairing gap.
To that end, we split the inverse propagator, \eq{Sinv},
into a free part and a gap dependent part, 
\beq
    S^{-1} = S_0^{-1} + \tilde{\Delta}\,,
\eeq
where $S_0=\left.S\right|_{\Delta=0}$ and expand the logarithm in
\eq{Om0} in a power series in $\tilde{\Delta}$: 
\begin{alignat}{1}
&\Tr\ln\left(\frac{1}{T}\left(S_0^{-1}+\tilde{\Delta}\right)\right)
\nonumber\\
=\;
&\Tr\ln\left(\frac{1}{T}S_0^{-1}\right)
-
\Tr\sum_{n=1}^\infty \frac{1}{n}\left(-S_0\,\tilde{\Delta}\right)^n
\,.
\end{alignat}
Since $S_0\,\tilde{\Delta}$ has only off-diagonal terms in Nambu-Gor'kov space,
only even $n$ contribute to the trace. 
To lowest nontrivial order, we thus get
\begin{alignat}{1}
\Omega_{MF}
\;=\quad
&\left.\Omega_{MF}\right|_{\Delta=0}
\nonumber \\
+\,&\frac{T}{4V}\Tr\left(S_0\tilde{\Delta}S_0\tilde{\Delta}\right)
\,+\,\frac{1}{4H}\sum_A\sum_{q_k} |\Delta_{A,q_k}|^2
\nonumber \\
+\,&O(\Delta^4)\,.
\label{OmOMGL}
\end{alignat}
For the evaluation of the trace, we need the explicit
expressions for $S_0$ and $\tilde{\Delta}$. 
From \eq{Sinv} we get
\beq
    S_0 = \left(\begin{array}{cc} S_0^+ & 0 \\ 0 & S_0^- \end{array}\right)
\quad \text{and} \quad
    \tilde\Delta = 
    \left(\begin{array}{cc} 0 & \tilde\Delta^+ \\ 
                            \tilde\Delta^- & 0 \end{array}\right)\,,
\eeq
with
\begin{alignat}{1}
    \left(\,S_0^\pm\,\right)_{p_m,p_n} &=\; 
    (\pslash_n \pm \muslash)^{-1}\,\delta_{p_m,p_n}
    \;\equiv\;S_0^\pm(p_n)\,\delta_{p_m,p_n}\,,
\nonumber \\
    \left(\tilde\Delta^+\right)_{p_m,p_n} &=\;
    \sum_{q_k}\hat\Delta_{q_k}\gamma_{5}\,\delta_{q_k,p_m-p_n}\,,
\end{alignat}
and $\tilde\Delta^- = - (\tilde\Delta^+)^\dagger$.
Furthermore, we get from \eq{Deltax} 
\beq
    \hat\Delta_{q_k} = \sum_A \Delta_{A,q_k} \,\tau_{A}\lambda_{A}\,,
\eeq
while $S_0$ is diagonal in flavor and color,
\beq
    S_0^\pm = \mathrm{diag}_{fc}\left((S_0^\pm)_{fc}\right)\,, 
\eeq
with flavor indices $f \in \{u,d,s\}$ and color indices $c \in \{r,g,b\}$.
Inserting these expressions into \eq{OmOMGL},
the trace is readily evaluated. 
As a consequence of momentum conservation, we find that unequal
Fourier components of the gap function do not interfere at this order.
Likewise, there is no interference between two components with different
color-flavor indices.
We thus obtain an incoherent sum over $|\Delta_{A,q_k}|^2$, 
which can be combined with the sum in \eq{OmOMGL} to get
\begin{alignat}{1}
\Omega_{MF}
\;=\quad
&\left.\Omega_{MF}\right|_{\Delta=0}
\nonumber \\
+\,&\frac{1}{2}\sum_A\sum_{q_k} \tilde\alpha_{A,q_k}\,
|\Delta_{A,q_k}|^2
\,+\,O(\Delta^4)\,.
\end{alignat}
The coefficients $\tilde\alpha_{A,q_k}$ are given by
\begin{alignat}{2}
 &\tilde\alpha_{2,q_k} \,= &&  
\nonumber\\
&-\frac{T}{2V} \sum_{p_n} \big(&
  &{\mathrm tr}_D\left[(S_0^+)_{ur}(p_n+q_k)\,\gamma_5\,
    (S_0^-)_{dg}(p_n)\,\gamma_5\right]
\nonumber\\[-2mm]
&& +\, &{\mathrm tr}_D\left[(S_0^+)_{ug}(p_n+q_k)\,\gamma_5\,
    (S_0^-)_{dr}(p_n)\,\gamma_5\right]
\nonumber\\[1mm]
&& +\, &{\mathrm tr}_D\left[(S_0^+)_{dr}(p_n+q_k)\,\gamma_5\,
    (S_0^-)_{ug}(p_n)\,\gamma_5\right]
\nonumber\\[1mm]
&& +\, &{\mathrm tr}_D\left[(S_0^+)_{dg}(p_n+q_k)\,\gamma_5\,
    (S_0^-)_{ur}(p_n)\,\gamma_5\right]\big)
\nonumber\\[1mm]
   &+\, \frac{1}{4H}\,,&&
\end{alignat}
and analogously for $\tilde\alpha_{5,q_k}$ and $\tilde\alpha_{7,q_k}$.
In the 2SC phase only $\Delta_{2,q_k}$ is nonvanishing, and we may drop 
the index $A$. 
The coefficients then essentially depend on the chemical potential 
difference $\delta\mu$, defined in \eq{deltamu}.

The remaining traces in Dirac space, denoted by ${\mathrm tr}_D$, are
trivial.
As before, the sum over $p_n$ should be read as a Matsubara sum and
a sum over the three-momentum $\vec p_n$, cf.~\eq{pn}. In the infinite
volume limit, the latter should be replaced by an integral. 
Moreover, we have to introduce a regularization procedure to render this
integral finite. 

As pointed out above, the quadratic term of the Ginzburg-Landau 
functional is an incoherent superposition of the contributions from the
different Fourier components. In particular, the coefficients 
$\tilde\alpha_{q_k}$  are completely independent of each other
and should therefore be equal to the coefficients in the 
Fulde-Ferrell phase at given $\vec q$ and $\delta\mu$.

The determination of $\tilde\alpha(\vec{q}_k,\delta\mu)$ in the 
Fulde-Ferrell case is a well-known exercise 
(see e.g.~Refs.~\cite{Alford:2000ze,Casalbuoni:2003wh}).
In a weak-coupling expansion at high densities one finds
\beq
\tilde\alpha(\vec{q}) =
\frac{\bar\mu^2}{\pi^2}
\big(\, \alpha(|\vec{q}|,\delta\mu) + \delta\alpha_{reg} \,\big)\,,
\eeq
where
\beq
\alpha(q,\delta\mu)
=
-1
+
\frac{\delta\mu}{2q}
\ln\left(\frac{q+\delta\mu}{q-\delta\mu}\right)
-
\frac{1}{2}
\ln\left(\frac{\Delta_{BCS}^{2}}{4\vert q-\delta\mu\,\vert^{2}}\right)
\,,
\eeq
while $\delta\alpha_{reg}$ depends explicitly on the regularization
in the sense that the cutoff dependence of this term cannot be
absorbed in an observable, like $\Delta_{BCS}$. 
As we pointed out in section \ref{sec:regularization}, 
such terms should be avoided in order to keep the results free from
regularization artifacts. 
In fact, in the regularization schemes usually employed for the 
Fulde-Ferrell phase \cite{Alford:2000ze,Casalbuoni:2003wh},
the dependence on the regularization can completely be absorbed into 
a $\Delta_{BCS}$ dependence, and $\delta\alpha_{reg}$ vanishes.

However, 
it is not
obvious whether this regularization can be generalized to arbitrary
gap functions in the thermodynamic potential.
Using a sharp cutoff $\Lambda$ for in- and out-going momenta instead, 
we obtain
\beq
\delta\alpha_{reg}^\mathit{cutoff} =
\frac{\vert\vec{q}\vert\,\Lambda^2}{4(\Lambda-\bar\mu)\bar\mu^2}
\,.
\eeq
This excludes this most naive regularization scheme.
For the proper-time regularization introduced in 
section \ref{sec:regularization}, on the other hand,
we find 
\beq
\delta\alpha_{reg}^\mathit{proper\; time} = 0
\eeq
in weak coupling. We therefore suggest to use this scheme for a
consistent regularization of inhomogeneous phases. 

Finally, we would like to add two comments:
First, instead of starting from \eq{Om0}, the Ginzburg-Landau functional
could be derived equally well by expanding \eq{Om0Mat}. 
At $T=0$ this means that
the Ginzburg-Landau expansion corresponds to a perturbative
expansion of the eigenvalues of the Hamiltonian,
with the pairing gap treated as perturbation. 
However, since some of the eigenvalues of the ``unperturbed'' Hamiltonian 
vanish at the Fermi surfaces of the particles, whereas the pairing gap 
is large, the perturbative expansion does actually not converge
for the most interesting eigenvalues near the Fermi surfaces.
As an example consider a BCS-like dispersion relation,
\begin{alignat}{1}
    E(p) &= \sqrt{(p-\mu)^2 + |\Delta|^2} 
\nonumber\\
    &= |p-\mu| + \frac{|\Delta|^2}{2|p-\mu|} + O(|\Delta|^4)\,,
\end{alignat}
which obviously does not converge near $p = \mu$.
The Ginzburg-Landau expansion may therefore suffer a similar problem.

Second, the Ginzburg-Landau expansion may be useful to identify the
divergencies in the thermodynamic potential.
In $3+1$ dimensions it can be used to show that all divergencies can be
absorbed by adding the expressions $c_2\int\vert\Delta(x)\vert^2$ and
$c_4\int\vert\Delta(x)\vert^4$ with appropriate coefficients $c_2$ and $c_4$
to the thermodynamic potential.

\section{Symmetry of the eigenvalue spectrum under $\delta\mu$}
\label{app:symevs}

In this appendix we discuss properties of the eigen\-spectrum of the
Hamiltonian ${\cal H}_{\Delta,\delta\mu}$ in \eq{HDm}.
For the case of a real gap function $\Delta(x)$ we would like to show, that in
terms of the eigenvalue spectrum
$\{E_\lambda\}$ of ${\cal H}_{\Delta,\delta\mu}$ the eigenvalue spectrum of 
${\cal H}_{\Delta,-\delta\mu}$ is given by $\{-E_\lambda\}$.
As a consequence, all eigenvalues of the total Hamiltonian
${\cal H}_\mathit{HDE}$, \eq{HHDE}, come in pairs $\{E_\lambda,-E_\lambda\}$.

Restricting to real condensates, we can write
\bea
{\cal H}_{\Delta,\delta\mu}
&=&
H_0\sigma_3+\Delta\sigma_1-\delta\mu \unity
\,,
\eea
where
$(H_0)_{\vec p_m,\vec p_n} \equiv (p_m - \bar\mu)\delta _{\vec p_m,\vec p_n}$,
$\Delta_{\vec p_m,\vec p_n} \equiv \Delta_{p_m - p_n}$ and $\{\sigma_i\}$ are
the conventional Pauli matrices. Here we used that the Fourier components
obey $\Delta_{q_k} = \Delta^*_{-q_k}$ for a real gap function, see
\eq{Deltaq}.
From
$\{\sigma_a,\sigma_b\}=2\delta_{ab}\unity$ we then obtain with $J=i\sigma_2$
\bea
{\cal H}_{\Delta,0}J
&=&
-J{\cal H}_{\Delta,0}
\,.
\eea
Consequently an eigenvector $v$ with eigenvalue $E$ will always come 
along with an eigenvector $Jv$ with eigenvalue $-E$, i.e., eigenvalues 
of ${\cal H}_{\Delta,0}$ come in pairs $(E_\lambda,-E_\lambda)$.

More specifically we can choose the eigensystem of $J$ as a basis in which
\bea
{\cal H}_{\Delta,0}
&=&
U
\left(\! \begin{array}{cc}
  0\; & -H_0+i\Delta \\  -H_0-i\Delta\; & 0
\end{array}\right)
U^\dagger
\,,
\eea
with
\bea
U
&=&
\frac{1}{\sqrt{2}}
\left(\! \begin{array}{rc}
  -i\; & i \\  1\; & 1
\end{array}\right)
\,.
\eea
Therefore we have
\bea
{\cal H}_{\Delta,0}^2
&=&
U
\left(\! \begin{array}{cc}
  H_0^2+V_+\; & 0 \\  0 & H_0^2+V_-
\end{array}\right)
U^\dagger
\,,
\eea
with
\bea
V_\pm
&=&
\Delta^2\pm i[H_0,\Delta]
\,.
\eea
One can easily show that the spectra of the two operators $H_0^2+V_\pm$
are identical and given by $\{E_\lambda^2\}$. This fact is sometimes 
referred to as being isospectral. The operators $V_\pm$ are furthermore
very similar to potentials of fermions and bosons in supersymmetric quantum
mechanics with superpotential $\Delta$.

Turning on again the chemical potential shift, we note that
\beq
    {\cal H}_{\Delta,\delta\mu} =  
    {\cal H}_{\Delta,0} - \delta\mu\,\unity~,
\eeq
i.e., $\delta\mu$ just leads to a shift of the eigenvalues
of ${\cal H}_{\Delta,0}$. 
Hence, for each pair $(E_\lambda,-E_\lambda)$ of the eigenvalues of 
${\cal H}_{\Delta,0}$, there is a corresponding pair
$(E_\lambda\mp\delta\mu,-E_\lambda\mp\delta\mu)$ of eigenvalues of
${\cal H}_{\Delta,\pm\delta\mu}$.  Therefore we get the claimed
connection between the spectrum of ${\cal H}_{\Delta,\delta\mu}$ and
${\cal H}_{\Delta,-\delta\mu}$.

Note that, in this proof, we made explicitly use of the fact that the matrix 
$\Delta$ is hermitian, which, in turn, was a consequence of our assumption
that the gap function $\Delta(x)$ is real. 
This is the case for the inhomogeneous solutions discussed in 
Sec.~\ref{sec:genreal} and thereafter, as well as for the BCS phase. 
It is not true for the FF phase, where the gap function is complex.
However, in this case we can explicitly see from \eq{FFeigenvalues}
that the eigenvalues $E_{\pm}(\vec k)$ go over into $-E_{\mp}(-\vec k)$ 
when we replace $\delta\mu$ by $-\delta\mu$.
Therefore, since we integrate over $\vec k$ in \eq{OmegaFF},
${\cal H}_{\Delta,\delta\mu}$ and ${\cal H}_{\Delta,-\delta\mu}$ contribute
equally to the thermodynamic potential.

\end{document}